\documentclass[11pt]{article} 

\usepackage{textcomp, amssymb}  
\usepackage{anysize}            
\usepackage{amsfonts}
\usepackage{amsmath}
\usepackage{dsfont}
\usepackage{MnSymbol}
\usepackage{mathtools}
\usepackage{graphicx}
\usepackage{epsfig}
\usepackage{epstopdf}
\usepackage{tikz}
\usepackage{amsthm}
\usepackage{ifpdf}
\usepackage{thm-restate}
\usepackage{stmaryrd}
\usepackage{cleveref}
\usepackage{subcaption}

\usetikzlibrary{calc,decorations.pathmorphing,shapes,graphs,positioning,automata,angles}

\newcounter{sarrow}

\newcounter{sarrow1}
\newcommand\xnrsquigarrow[1]{%
\stepcounter{sarrow1}%
\mathrel{\begin{tikzpicture}[baseline= {( $ (current bounding box.south) + (0,-0.5ex) $ )}]
\node[inner sep=.5ex] (\thesarrow) {$\scriptstyle #1$};
\path[draw,<-,decorate,
  decoration={zigzag,amplitude=0.7pt,segment length=1.2mm,pre=lineto,pre length=4pt}]
    (\thesarrow1.south east) -- (\thesarrow1.south west);
    $\slashedarrowfill@\relbar\relbar/$
    \end{tikzpicture}}%
}

\makeatletter
\def\slashedarrowfill@#1#2#3#4#5{%
  $\m@th\thickmuskip0mu\medmuskip\thickmuskip\thinmuskip\thickmuskip
   \relax#5#1\mkern-7mu%
   \cleaders\hbox{$#5\mkern-2mu#2\mkern-2mu$}\hfill
   \mathclap{#3}\mathclap{#2}%
   \cleaders\hbox{$#5\mkern-2mu#2\mkern-2mu$}\hfill
   \mkern-7mu#4$%
}
\def\rightslashedarrowfillb@{%
  \slashedarrowfill@\relbar\relbar/\rightarrow}
\newcommand\xnrightarrow[2][]{%
  \ext@arrow 0055{\rightslashedarrowfillb@}{#1}{#2}}

\def\rightslashedarrowfille@{%
  \slashedarrowfill@\relbar\relbar/\twoheadrightarrow}
\newcommand\xntworightarrow[2][]{%
  \ext@arrow 0055{\rightslashedarrowfille@}{#1}{#2}}

\def\rightslashedarrowfillg@{%
  \slashedarrowfill@\relbar\relbar{\raisebox{.12em}{}}\twoheadrightarrow}
\newcommand\xtworightarrow[2][]{%
  \ext@arrow 0055{\rightslashedarrowfillg@}{#1}{#2}}

\def\rightslashedarrowfillx@{%
  \slashedarrowfill@\Relbar\Relbar/\rightrightarrows}
\newcommand\xnTworightarrow[2][]{%
  \ext@arrow 0055{\rightslashedarrowfillx@}{#1}{#2}}

\def\rightslashedarrowfilly@{%
  \slashedarrowfill@\Relbar\Relbar{\raisebox{.12em}{}}\rightrightarrows}
\newcommand\xTworightarrow[2][]{%
  \ext@arrow 0055{\rightslashedarrowfilly@}{#1}{#2}}

\pgfdeclareshape{slash underlined}
{
  \inheritsavedanchors[from=rectangle] 
  \inheritanchorborder[from=rectangle]
  \inheritanchor[from=rectangle]{north}
  \inheritanchor[from=rectangle]{north west}
  \inheritanchor[from=rectangle]{north east}
  \inheritanchor[from=rectangle]{center}
  \inheritanchor[from=rectangle]{west}
  \inheritanchor[from=rectangle]{east}
  \inheritanchor[from=rectangle]{mid}
  \inheritanchor[from=rectangle]{mid west}
  \inheritanchor[from=rectangle]{mid east}
  \inheritanchor[from=rectangle]{base}
  \inheritanchor[from=rectangle]{base west}
  \inheritanchor[from=rectangle]{base east}
  \inheritanchor[from=rectangle]{south}
  \inheritanchor[from=rectangle]{south west}
  \inheritanchor[from=rectangle]{south east}
  \inheritanchorborder[from=rectangle]
  \foregroundpath{
    \southwest \pgf@xa=\pgf@x \pgf@ya=\pgf@y
    \northeast \pgf@xb=\pgf@x \pgf@yb=\pgf@y
    \pgf@xc=\pgf@xa
    \advance\pgf@xc by .5\pgf@xb
    \pgf@yc=\pgf@ya 
    \advance\pgf@xc by -1.3pt
    \advance\pgf@yc by -1.8pt
    \pgfpathmoveto{\pgfqpoint{\pgf@xc}{\pgf@yc}}
    \advance\pgf@xc by  2.6pt
    \advance\pgf@yc by  3.6pt
    \pgfpathlineto{\pgfqpoint{\pgf@xc}{\pgf@yc}}
    \pgfpathmoveto{\pgfqpoint{\pgf@xa}{\pgf@ya}}
    \pgfpathlineto{\pgfqpoint{\pgf@xb}{\pgf@ya}}
 }
}
\tikzset{nomorepostaction/.code=\let\tikz@postactions\pgfutil@empty}

\newcommand{\semangle}[1]{\langle\!|#1|\!\rangle}
\newcommand{\sembrack}[1]{\llbracket #1\rrbracket}
\newcommand*{\rmbrace}{|\mskip-4mu\}}
\newcommand*{\lmbrace}{\{\mskip-4mu|}
\newcommand*{\mset}[1]{\lmbrace#1\rmbrace}
\newcommand*{\rsbrace}{|\mskip-4mu\rangle}
\newcommand*{\lsbrace}{\langle\mskip-4mu|}
\newcommand*{\step}[1]{\lsbrace#1\rsbrace}

\newtheorem{theorem}{Theorem}[section]
\newtheorem{definition}[theorem]{Definition}
\newtheorem{proposition}[theorem]{Proposition}
\newtheorem{lemma}[theorem]{Lemma}

\newtheorem{corollary}[theorem]{Corollary}
\newcommand{\pretesting}{\mathrel{\ooalign{\raise0.13ex\hbox{$\sqsubset$}\cr\hidewidth\raise-0.6ex\hbox{\scalebox{0.9}{$\sim$}}\hidewidth\cr}}}
\newcommand{\prepomset}{\lesssim_{\mathrm{P}}}

\newcommand{\prepomsets}[1]{\lesssim_{\mathrm{(P,#1)}}}

\newcommand{\preeq}{{\mathrel{\ooalign{\raise0.13ex\hbox{$=$}\cr\hidewidth\raise-0.6ex\hbox{\scalebox{0.9}{$\sim$}}\hidewidth\cr}}}}

\newcommand{\fix}{\mathrm{fix}}
\begin{document}

\begin{titlepage}
\thispagestyle{empty}

\hrule
\begin{center}
{\bf\LARGE Structural Operational Semantics for True Concurrency\\}
%
\vspace{0.5cm}
--- Yong Wang ---

\vspace{2cm}

\end{center}
\end{titlepage}

\newpage 

\setcounter{page}{1}\pagenumbering{roman}

\tableofcontents

\newpage
\setcounter{page}{1}\pagenumbering{arabic}

        \section{Introduction}\label{intro}

Structural operational semantics (SOS) \cite{SOS} surveyed the foundations on operational semantics for process theory. There existed many works on these foundations, and we do not repeatedly enumerate all the works and just refer to \cite{SOS} as the origin. Mainly, SOS \cite{SOS} was based on the interleaving semantics, i.e., a transition occurs by the way of executing just one single action. SOS \cite{SOS} provides the concepts of Labelled Transition System (LTS), Transition System Specification (TSS), enumeration of interleaving behavioural equivalences, conservative extension, and also deep insights on these concepts, including the meanings of TSSs, congruence formats of TSSs, higher-order languages and denotational semantics.

Traditionally, in concurrency theory, the so-called true concurrency is usually defined by causal order graphs, such as event structures \cite{ES} \cite{IES}, Petri nets \cite{PN01} \cite{PN02}, etc.

The works on truly concurrent process algebras \cite{APTC} \cite{APTC2} bridge interleaving concurrency and true concurrency. Truly concurrent process algebras are generalizations of the corresponding process algebras from interleaving concurrency to true concurrency, such as Calculus for True concurrency (CTC) \cite{APTC} to Milner's CCS \cite{CCS}, Algebra of parallelism for True Concurrency (APTC) \cite{APTC} to Bergstra and Klop's ACP \cite{ACP}, mobile calculus $\pi_{tc}$ \cite{APTC} to Milner's $\pi$ calculus \cite{PI1} \cite{PI2}.

It is natural that we can extend SOS \cite{SOS} to SOS for true concurrency. From SOS \cite{SOS} to SOS for true concurrency, it is in nature to give the related concepts in SOS \cite{SOS} a truly concurrent semantics foundation, i.e., a transition occurs by executing a Partially Ordered Multi Set (pomset) of actions replacing just one single action. Under the framework of SOS \cite{SOS}, for the extension to the truly concurrent one, something are changing: LTS is generalized to Pomset LTS (PLTS), TSS to Pomset TSS (PTSS), interleaving behavioural equivalences to truly concurrent ones, congruence formats of TSSs to those of PTSSs; something are remained, such as the concept of conservative extension, the meanings of TSSs and PTSSs. 

Originally, we had planned to give a complete truly concurrent version of SOS according to the framework of SOS \cite{SOS}, two reasons: (1) something from SOS \cite{SOS} to truly concurrent one remain unchanged; (2) some works for true concurrency are still absent, for example, some truly concurrent process algebras, such as SCCS \cite{SCCS} for true concurrency, Meije \cite{Meije} for true concurrency, lead to this version of SOS for true concurrency. In this version, we write some conclusions or conjectures without any proofs. Indeed, a full version of SOS for true concurrency still needs plenty of work, and let it be a future one. 
\newpage\section{Pomset Labelled Transition Systems}\label{plts} 

In this chapter, we introduce pomset and pomset labelled transition systems in \cref{plts}. In \cref{tcbe}, we introduce truly concurrent behavioural equivalences, including some bisimulation equivalences and pomset trace equivalences. In \cref{bcl}, we introduce Baldan-Crafa logic and its fragments which are versions of Hennessy-Milner logic modulo truly concurrent bisimilarities. Then we introduce the backgrounds of term algebras in \cref{ta} and pomset transition system specifications (PTSS) in \cref{ptss}. Finally, in \cref{eptss}, we introduce some examples of PTSSs, including PTSSs for Basic Process Algebra with Empty Process, parallelism, priority and discrete time.

\subsection{Pomset Labelled Transition Systems}\label{plts}

Assume a fixed but arbitrary set of actions $\mathsf{Act}$. 

\begin{definition}[Poset of actions]
A poset of actions with partial order among them is a pair $\mathbf{u}=(A, \leq)$, where $A\subseteq\mathsf{Act}$ is the carrier set, $\leq$ is a partial order on $A$.

For a poset $\mathbf{u}$, $A_{\mathbf{u}}$ and $\leq_{\mathbf{u}}$ denote the carrier and the partial order of $\mathbf{u}$ respectively. The set of posets is denoted $\mathsf{Pos}$ and the empty poset is $\mathbf{\epsilon}$.
\end{definition}

\begin{definition}[Poset morphism]
For posets $(A,\leq)$ and $(A',\leq')$ and function $f:A\rightarrow A'$, $f$ is called a poset morphism if for $a_0,a_1\in A$ with $a_0\leq a_1$, then $f(a_0)\leq' f(a_1)$ holds.
\end{definition}

\begin{definition}[Poset isomorphism]
Let $\mathbf{u}=(A_1,\leq_1)$ and $\mathbf{v}=(A_2,\leq_2)$ be posets. A poset morphism $h$ from $\mathbf{u}=(A_1,\leq_1)$ to $\mathbf{v}=(A_2,\leq_2)$ is a poset morphism from $(A_1,\leq_1)$ and $(A_2,\leq_2)$. Moreover, $h$ is a poset isomorphism if it is a bijection with $h^{-1}$ is a poset isomorphism from $(A_2,\leq_2)$ to $(A_1,\leq_1)$. We say that $\mathbf{u}=(A_1,\leq_1)$ is isomorphic to $\mathbf{v}=(A_2,\leq_2)$ denoted $(A_1,\leq_1)\sim(A_2,\leq_2)$, if there exists a poset isomorphism $h$ between $(A_1,\leq_1)$ and $(A_2,\leq_2)$.
\end{definition}

It is easy to see that $\sim$ is an equivalence and can be used to abstract from the carriers.

\begin{definition}[Pomset]
A partially ordered multiset, pomset, is a $\sim$-equivalence class of posets. The $\sim$-equivalence class of $\mathbf{u}\in\mathsf{Pos}$ is denoted $[\mathbf{u}]$; the set of pomsets is denoted $\mathsf{Pom}$; the empty poset is denoted $\mathbf{\epsilon}$ and the $\sim$-equivalence class of $\mathbf{\epsilon}$ is also denoted by $\epsilon$; the pomset containing exactly one action $a\in\mathsf{Act}$ is called primitive.
\end{definition}

\begin{definition}[Pomset composition in parallel]
Let $U,V\in\mathsf{Pom}$ with $U=[\mathbf{u}]$ and $V=[\mathbf{v}]$. We write $U\parallel V$ for the parallel composition of $U$ and $V$, which is the pomset represented by $\mathbf{u}\parallel\mathbf{v}$, where

$$A_{\mathbf{u}\parallel\mathbf{v}}=A_{\mathbf{u}}\cup A_{\mathbf{v}}
\quad\quad\leq_{\mathbf{u}\parallel\mathbf{v}}=\leq_{\mathbf{u}}\cup\leq_{\mathbf{v}}$$
\end{definition}

\begin{definition}[Pomset composition in sequence]
Let $U,V\in\mathsf{Pom}$ with $U=[\mathbf{u}]$ and $V=[\mathbf{v}]$. We write $U\cdot V$ for the sequential composition of $U$ and $V$, which is the pomset represented by $\mathbf{u}\cdot\mathbf{v}$, where

$$A_{\mathbf{u}\cdot\mathbf{v}}=A_{\mathbf{u}}\cup A_{\mathbf{v}}
\quad\quad\leq_{\mathbf{u}\cdot\mathbf{v}}=\leq_{\mathbf{u}}\cup\leq_{\mathbf{v}}\cup(A_{\mathbf{u}}\times A_{\mathbf{v}})$$
\end{definition}

\begin{definition}[Pomset types]
Let $U\in\mathsf{Pom}$, $U$ is sequential (resp. parallel) if there exist non-empty pomsets $U_1$ and $U_2$ such that $U=U_1\cdot U_2$ (resp. $U=U_1\parallel U_2$).
\end{definition}

\begin{definition}[Factorization]
Let $U\in\mathsf{Pom}$. (1) When $U=U_1\cdot\cdots \cdot U_i\cdot\cdots \cdot U_n$ with each $U_i$ non-sequential and non-empty, the sequence $U_1,\cdots,U_i,\cdots,U_n$ is called a sequential factorization of $U$. (2) When $U=U_1\parallel\cdots \parallel U_i\parallel\cdots \parallel U_n$ with each $U_i$ non-parallel and non-empty, the multiset $\mset{U_1,\cdots,U_i,\cdots,U_n}$ is called a parallel factorization of $U$.
\end{definition}

\begin{lemma}[Factorization]\label{LemmaFactorization}
Sequential and parallel factorizations exist uniquely.
\end{lemma}

\begin{restatable}[Pomset labelled transition system]{definition}{pomsetlts}\label{pomlts}
A pomset labelled transition system (PLTS) is a quadruple $(\mathsf{Proc},\mathsf{Act},\{\xrightarrow{U}|U\in\mathsf{Pom}\},\mathsf{Pred})$, where:

\begin{enumerate}
  \item $\mathsf{Proc}$ is a set of states, ranged over by $s,s'$.
  \item $\mathsf{Act}$ is a set of actions, ranged over by $a,b$.
  \item $\mathsf{Pom}$ is the set of pomsets over $\mathsf{Act}$, ranged over by $U,V$.
  \item $\xrightarrow{U}\subseteq\mathsf{Proc}\times\mathsf{Proc}$ is called a pomset transition for every $U\in\mathsf{Pom}$. We write $s\xrightarrow{U}s'$ instead of $(s,s')\in\xrightarrow{U}$, and write $s\xnrightarrow{U}$ if $s\xrightarrow{U}s'$ with no state $s'$. Intuitively, $s\xrightarrow{U}s'$ means that state $s$ can evolve into state $s'$ by the execution of pomset $U$. We see that traditional single action transition $s\xrightarrow{a}s'$ with $a\in\mathsf{Act}$ is a special case of pomset transition in which the pomset is primitive.
  \item $P\subseteq\mathsf{Proc}$ for every $P\in\mathsf{Pred}$. We write $sP$ (resp. $s\neg P$) if state $s$ satisfies (resp. does not satisfy) predicate $P$. Intuitively, $sP$ means that predicate $P$ holds in state $s$.
\end{enumerate}

The binary pomset transitions $s\xrightarrow{U}s'$ and unary predicates $sP$ in a PLTS are called transitions.
\end{restatable}

\begin{definition}[Finiteness conditions on a PLTS]
The finiteness conditions on a PLTS is similar to a labelled transition system (LTS), a PLTS is:

\begin{itemize}
  \item Finitely branching: if for every state $s$ there are only finitely many outgoing pomset transitions $s\xrightarrow{U}s'$.
  \item Regular: if it is finitely branching and each state can reach only finitely many other states.
  \item Finite: if it is finitely branching and there is no infinite sequence of pomset transitions $s_0\xrightarrow{U_0}s_1\xrightarrow{U_1}\cdots$.
\end{itemize}
\end{definition}

\subsection{Truly Concurrent Behaviourial Equivalences}\label{tcbe}

\begin{definition}[Configuration]
Let $\mathcal{P}$ be a PLTS. A (finite) configuration in $\mathcal{P}$ is a (finite) sub-pomset of $\mathsf{Act}(\mathcal{P})$ which denotes $\mathsf{Act}$ of $\mathcal{P}$, $\mathbf{C}\subseteq \mathsf{Act}(\mathcal{P})$. The set of finite configurations of $\mathcal{P}$ is denoted by $\mathcal{C}(\mathcal{P})$. It is obvious that there is an one-to-one correspondence between the configurations of $\mathcal{P}$ and the states $\mathsf{Proc}$ of $\mathcal{P}$.
\end{definition}

\begin{definition}[Pomset transitions and step of configuration]
Let $\mathcal{P}$ be a PLTS and let $\mathbf{C}\in\mathcal{C}(\mathcal{P})$, and $\emptyset\neq U\subseteq \mathsf{Act}(\mathcal{P})$, if $\mathbf{C}\cap U=\emptyset$ and $\mathbf{C}'=\mathbf{C}\cup U\in\mathcal{C}(\mathcal{P})$, then $\mathbf{C}\xrightarrow{U} \mathbf{C}'$ is called a pomset transition from $\mathbf{C}$ to $\mathbf{C}'$. When the events in $U$ are pairwise concurrent (with no partial orders pairwise), we say that $\mathbf{C}\xrightarrow{U}\mathbf{C}'$ is a step. It is obvious that there is an one-to-one correspondence between pomset transition defined by the configurations of $\mathcal{P}$ and the states $\mathsf{Proc}$ inner $\mathcal{P}$.
\end{definition}

\begin{definition}[Pomset, step bisimulation]\label{PSB}
Let $\mathcal{P}$ be a PLTS. A pomset bisimulation is a relation $R\subseteq\mathcal{C}(\mathcal{P})\times\mathcal{C}(\mathcal{P})$, such that: (1) if $(\mathbf{C}_1,\mathbf{C}_2)\in R$, and $\mathbf{C}_1\xrightarrow{U_1}\mathbf{C}_1'$ then $\mathbf{C}_2\xrightarrow{U_2}\mathbf{C}_2'$, with $U_1\subseteq \mathsf{Act}(\mathcal{P})$, $U_2\subseteq \mathsf{Act}(\mathcal{P})$, $U_1\sim U_2$ and $(\mathbf{C}_1',\mathbf{C}_2')\in R$, and vice-versa; (2) if $(\mathbf{C}_1,\mathbf{C}_2)\in R$, and $\mathbf{C}_1P$ then $\mathbf{C}_2P$, and vice-versa. We say that $\mathbf{C}_1$, $\mathbf{C}_2$ are pomset bisimilar, written $\mathbf{C}_1\sim_p \mathbf{C}_2$, if there exists a pomset bisimulation $R$, such that $(\emptyset,\emptyset)\in R$. By replacing pomset transitions with steps, we can get the definition of step bisimulation. When $\mathbf{C}_1$ and $\mathbf{C}_2$ are step bisimilar, we write $\mathbf{C}_1\sim_s \mathbf{C}_2$.
\end{definition}

\begin{definition}[Pomset, step simulation]\label{PSS}
Let $\mathcal{P}$ be a PLTS. A pomset simulation is a relation $R\subseteq\mathcal{C}(\mathcal{P})\times\mathcal{C}(\mathcal{P})$, such that: (1) if $(\mathbf{C}_1,\mathbf{C}_2)\in R$, and $\mathbf{C}_1\xrightarrow{U_1}\mathbf{C}_1'$ then $\mathbf{C}_2\xrightarrow{U_2}\mathbf{C}_2'$, with $U_1\subseteq \mathsf{Act}(\mathcal{P})$, $U_2\subseteq \mathsf{Act}(\mathcal{P})$, $U_1\sim U_2$ and $(\mathbf{C}_1',\mathbf{C}_2')\in R$; (2) if $(\mathbf{C}_1,\mathbf{C}_2)\in R$, and $\mathbf{C}_1P$ then $\mathbf{C}_2P$. We say that $\mathbf{C}_1$, $\mathbf{C}_2$ are pomset similar, written $\mathbf{C}_1\lesssim_p \mathbf{C}_2$, if there exists a pomset simulation $R$, such that $(\emptyset,\emptyset)\in R$. By replacing pomset transitions with steps, we can get the definition of step simulation. When $\mathbf{C}_1$ and $\mathbf{C}_2$ are step similar, we write $\mathbf{C}_1\lesssim_s \mathbf{C}_2$.
\end{definition}

\begin{definition}[Ready pomset, step simulation]\label{RPSS}
Let $\mathcal{P}$ be a PLTS. A ready pomset simulation is a relation $R\subseteq\mathcal{C}(\mathcal{P})\times\mathcal{C}(\mathcal{P})$, such that: (1) if $(\mathbf{C}_1,\mathbf{C}_2)\in R$, and $\mathbf{C}_1\xrightarrow{U_1}\mathbf{C}_1'$ then $\mathbf{C}_2\xrightarrow{U_2}\mathbf{C}_2'$, with $U_1\subseteq \mathsf{Act}(\mathcal{P})$, $U_2\subseteq \mathsf{Act}(\mathcal{P})$, $U_1\sim U_2$ and $(\mathbf{C}_1',\mathbf{C}_2')\in R$; (2) if $(\mathbf{C}_1,\mathbf{C}_2)\in R$, and $\mathbf{C}_1P$ then $\mathbf{C}_2P$; (3) if $(\mathbf{C}_1,\mathbf{C}_2)\in R$, and $\mathbf{C}_1\xnrightarrow{U_1}$ then $\mathbf{C}_2\xnrightarrow{U_2}$, with $U_1\subseteq \mathsf{Act}(\mathcal{P})$, $U_2\subseteq \mathsf{Act}(\mathcal{P})$ and $U_1\sim U_2$; (4) if $(\mathbf{C}_1,\mathbf{C}_2)\in R$, and $\mathbf{C}_1\neg P$ then $\mathbf{C}_2\neg P$. We say that $\mathbf{C}_1$, $\mathbf{C}_2$ are ready pomset similar, written $\mathbf{C}_1\lesssim_{rp} \mathbf{C}_2$, if there exists a ready pomset simulation $R$, such that $(\emptyset,\emptyset)\in R$. By replacing pomset transitions with steps, we can get the definition of ready step simulation. When $\mathbf{C}_1$ and $\mathbf{C}_2$ are ready step similar, we write $\mathbf{C}_1\lesssim_{rs} \mathbf{C}_2$.
\end{definition}

\begin{definition}[Posetal product]
Given a PLTS $\mathcal{P}$, the posetal product of their configurations, denoted $\mathcal{C}(\mathcal{P})\overline{\times}\mathcal{C}(\mathcal{P})$, is defined as

$$\{(\mathbf{C}_1,f,\mathbf{C}_2)|\mathbf{C}_1\in\mathcal{C}(\mathcal{P}),\mathbf{C}_2\in\mathcal{C}(\mathcal{P}),f:\mathbf{C}_1\rightarrow \mathbf{C}_2 \textrm{ isomorphism}\}$$

A subset $R\subseteq\mathcal{C}(\mathcal{P})\overline{\times}\mathcal{C}(\mathcal{P})$ is called a posetal relation. We say that $R$ is downward closed when for any $(\mathbf{C}_1,f,\mathbf{C}_2),(\mathbf{C}_1',f',\mathbf{C}_2')\in \mathcal{C}(\mathcal{P})\overline{\times}\mathcal{C}(\mathcal{P})$, if $(\mathbf{C}_1,f,\mathbf{C}_2)\subseteq (\mathbf{C}_1',f',\mathbf{C}_2')$ pointwise and $(\mathbf{C}_1',f',\mathbf{C}_2')\in R$, then $(\mathbf{C}_1,f,\mathbf{C}_2)\in R$.

For $f:U_1\rightarrow U_2$, we define $f[a_1\mapsto a_2]:U_1\cup\{a_1\}\rightarrow U_2\cup\{a_2\}$, $z\in U_1\cup\{a_1\}$,(1)$f[a_1\mapsto a_2](z)=
a_2$,if $z=a_1$;(2)$f[a_1\mapsto a_2](z)=f(z)$, otherwise. Where $U_1\subseteq \mathsf{Act}(\mathcal{P})$, $U_2\subseteq \mathsf{Act}(\mathcal{P})$, $a_1\in \mathsf{Act}(\mathcal{P})$, $a_2\in \mathsf{Act}(\mathcal{P})$.
\end{definition}

\begin{definition}[(Hereditary) history-preserving bisimulation]\label{HHPB}
A history-preserving (hp-) bisimulation is a posetal relation $R\subseteq\mathcal{C}(\mathcal{P})\overline{\times}\mathcal{C}(\mathcal{P})$ such that: (1) if $(\mathbf{C}_1,f,\mathbf{C}_2)\in R$, and $\mathbf{C}_1\xrightarrow{a_1} \mathbf{C}_1'$, then $\mathbf{C}_2\xrightarrow{a_2} \mathbf{C}_2'$, with $(\mathbf{C}_1',f[a_1\mapsto a_2],\mathbf{C}_2')\in R$, and vice-versa; (2) if $(\mathbf{C}_1,f,\mathbf{C}_2)\in R$, and $\mathbf{C}_1P$ then $\mathbf{C}_2P$, and vice-versa. $\mathbf{C}_1,\mathbf{C}_2$ are history-preserving (hp-)bisimilar and are written $\mathbf{C}_1\sim_{hp}\mathbf{C}_2$ if there exists an hp-bisimulation $R$ such that $(\emptyset,\emptyset,\emptyset)\in R$.

A hereditary history-preserving (hhp-)bisimulation is a downward closed hp-bisimulation. $\mathbf{C}_1,\mathbf{C}_2$ are hereditary history-preserving (hhp-)bisimilar and are written $\mathbf{C}_1\sim_{hhp}\mathbf{C}_2$.
\end{definition}

\begin{definition}[(Hereditary) history-preserving simulation]\label{HHPS}
A history-preserving (hp-) simulation is a posetal relation $R\subseteq\mathcal{C}(\mathcal{P})\overline{\times}\mathcal{C}(\mathcal{P})$ such that: (1) if $(\mathbf{C}_1,f,\mathbf{C}_2)\in R$, and $\mathbf{C}_1\xrightarrow{a_1} \mathbf{C}_1'$, then $\mathbf{C}_2\xrightarrow{a_2} \mathbf{C}_2'$, with $(\mathbf{C}_1',f[a_1\mapsto a_2],\mathbf{C}_2')\in R$; (2) if $(\mathbf{C}_1,f,\mathbf{C}_2)\in R$, and $\mathbf{C}_1P$ then $\mathbf{C}_2P$. $\mathbf{C}_1,\mathbf{C}_2$ are history-preserving (hp-)similar and are written $\mathbf{C}_1\lesssim_{hp}\mathbf{C}_2$ if there exists an hp-simulation $R$ such that $(\emptyset,\emptyset,\emptyset)\in R$.

A hereditary history-preserving (hhp-)simulation is a downward closed hp-simulation. $\mathbf{C}_1,\mathbf{C}_2$ are hereditary history-preserving (hhp-)similar and are written $\mathbf{C}_1\lesssim_{hhp}\mathbf{C}_2$.
\end{definition}

\begin{definition}[Ready (hereditary) history-preserving simulation]\label{RHHPS}
A ready history-preserving (hp-) simulation is a posetal relation $R\subseteq\mathcal{C}(\mathcal{P})\overline{\times}\mathcal{C}(\mathcal{P})$ such that: (1) if $(\mathbf{C}_1,f,\mathbf{C}_2)\in R$, and $\mathbf{C}_1\xrightarrow{a_1} \mathbf{C}_1'$, then $\mathbf{C}_2\xrightarrow{a_2} \mathbf{C}_2'$, with $(\mathbf{C}_1',f[a_1\mapsto a_2],\mathbf{C}_2')\in R$; (2) if $(\mathbf{C}_1,f,\mathbf{C}_2)\in R$, and $\mathbf{C}_1P$ then $\mathbf{C}_2P$; (3) if $(\mathbf{C}_1,f,\mathbf{C}_2)\in R$, and $\mathbf{C}_1\xnrightarrow{a}$, then $\mathbf{C}_2\xnrightarrow{a}$; (4) if $(\mathbf{C}_1,f,\mathbf{C}_2)\in R$, and $\mathbf{C}_1\neg P$ then $\mathbf{C}_2\neg P$. $\mathbf{C}_1,\mathbf{C}_2$ are ready history-preserving (hp-)similar and are written $\mathbf{C}_1\lesssim_{rhp}\mathbf{C}_2$ if there exists a ready hp-simulation $R$ such that $(\emptyset,\emptyset,\emptyset)\in R$.

A ready hereditary history-preserving (hhp-)simulation is a downward closed ready hp-simulation. $\mathbf{C}_1,\mathbf{C}_2$ are ready hereditary history-preserving (hhp-)similar and are written $\mathbf{C}_1\lesssim_{rhhp}\mathbf{C}_2$.
\end{definition}

\begin{definition}[Pomset trace semantics]
For a PLTS, a trace is a sequence of symbols in $\mathsf{Act}$, while a pomset trace is a trace within which each symbol is replaced by a pomset in $\mathsf{Pom}$.

$$\varsigma=\mset{a_{11},\cdots,a_{1m_1}}\cdots\mset{a_{n1},\cdots,a_{nm_n}}\in\mathsf{Pom}^*$$

for $m,n\in\mathbb{N}$ and $s_0\in\mathsf{Proc}$, if there exist states $s_1,\cdots,s_n\in\mathsf{Proc}$ such that $s_0\xrightarrow{\mset{a_{11},\cdots,a_{1m_1}}}s_1\xrightarrow{\mset{a_{21},\cdots,a_{2m_2}}}\cdots\xrightarrow{\mset{a_{n1},\cdots,a_{nm_n}}}s_n$, abbreviated by $s_0\xrightarrow{\varsigma}s_n$. For $P\in\mathsf{Pred}$, $\varsigma\in\mathsf{Pom}^*$ and $s,s'\in\mathsf{Proc}$, we write $\varsigma P$ if $s\xrightarrow{\varsigma}s' P$. And we write $s\lesssim_{t}s'$ if the set of pomset traces of $s$ is included in that of $s'$.
\end{definition}

For $s\in\mathsf{Proc}$, we define the following set:

$$\mathsf{initial}(s)\triangleq\{U\in\mathsf{Pom}|\exists s'\in\mathsf{Proc}~s\xrightarrow{U}s'\}\cup\{P\in\mathsf{Pred}|sP\}$$

\begin{definition}[Decorated pomset trace semantics]
Assume a PLTS.

\begin{itemize}
  \item Ready pomset traces. For $m,n\in\mathbb{N}$, a pomset sequence $X_0\mset{a_{11},\cdots,a_{1m_1}}X_1\cdots\mset{a_{n1},\cdots,a_{nm_n}}X_n$ is a ready pomset trace of state $s_0$ if $s_0\xrightarrow{\mset{a_{11},\cdots,a_{1m_1}}}s_1\xrightarrow{\mset{a_{21},\cdots,a_{2m_2}}}\cdots\xrightarrow{\mset{a_{n1},\cdots,a_{nm_n}}}s_n$ and $\mathsf{initial}(s_i)=X_i$, where $X_i\subseteq\mathsf{Pom}\cup\mathsf{Pred}$, $\mset{a_{i1},\cdots,a_{im_i}}\in\mathsf{Pom}$ and $i=0,\cdots,n$. We write $s\lesssim_{rt}s'$ if the set of ready pomset traces of $s$ is included in that of $s'$.
  \item Failure pomset traces. For $X\subseteq\mathsf{Pom}\cup\mathsf{Pred}$, we define $s_0\xrightarrow{X}s_1$ if $s_0=s_1$ and $\mathsf{initial}(s_0)\cap X=\emptyset$. A sequence $\varsigma\in(\mathsf{Pom}\cup2^{\mathsf{Pom}\cup\mathsf{Pred}})^*$ is a failure pomset trace of state $s_0$ if $s_0\xrightarrow{\varsigma}s_1$ for some state $s_1$. We write $s\lesssim_{ft}s'$ if the set of failure pomset traces of $s$ is included in that of $s'$.
  \item Readies. A pair $(\varsigma,X)$ with $\varsigma\in\mathsf{Pom}^*$ and $X\subseteq\mathsf{Pom}\cup\mathsf{Pred}$ is a ready of state $s_0$ if $s_0\xrightarrow{\varsigma}s_1$ for some state $s_1$ with $\mathsf{initial}(s_1)=X$. We write $s\lesssim_{r}s'$ if the set of readies of $s$ is included in that of $s'$.
  \item Failures. A pair $(\varsigma,X)$ with $\varsigma\in\mathsf{Pom}^*$ and $X\subseteq\mathsf{Pom}\cup\mathsf{Pred}$ is a failure of state $s_0$ if $s_0\xrightarrow{\varsigma}s_1$ for some state $s_1$ with $\mathsf{initial}(s_1)\cap X=\emptyset$. We write $s\lesssim_{f}s'$ if the set of failures of $s$ is included in that of $s'$.
  \item Completed pomset traces. $\varsigma\in\mathsf{Pom}^*$ is a completed pomset trace of state $s_0$ if $s_0\xrightarrow{\varsigma}s_1$ for some state $s_1$ with $\mathsf{initial}(s_1)=\emptyset$. Moreover, $\varsigma P$ with $\varsigma\in\mathsf{Pom}^*$ and $P\in\mathsf{Pred}$ is a completed pomset trace of state $s_0$ if $s_0\xrightarrow{\varsigma}s_1 P$ for some state $s_1$. We write $s\lesssim_{ct}s'$ if the set of completed pomset traces of $s$ is included in that of $s'$.
  \item Accepting pomset traces. let $\surd\in\mathsf{Pred}$ of a PLTS. $\varsigma\in\mathsf{Pom}^*$ is an accepting pomset trace of state $s_0$ if $s_0\xrightarrow{\varsigma}s_1\surd$ for some state $s_1$. We write $s\lesssim_{at}s'$ if the set of accepting pomset traces of $s$ is included in that of $s'$.
\end{itemize}
\end{definition}

For $\Theta\in\{p,s,hp,hhp,rp,rs,rhp,rhhp,ct,rt,ft,f,r,at,t\}$, the relation $\lesssim_{\Theta}$ is a preorder over states in arbitrary PLTSs. Its kernel is denoted by $\sim_{\Theta}$, i.e., $s\sim_{\Theta}s'$ if and only if $s\lesssim_{\Theta}s'$ and $s'\lesssim_{\Theta}s$. Then we can get the following conclusion.

\begin{proposition}
In any PLTS, the inclusions of preorders in $\Theta$ is shown in \cref{IOP}, where $\rightarrow$ from one relation to another means that the source of $\rightarrow$ is included in the target. 

\begin{figure}
  \centering
  \begin{tikzpicture}[every node/.style={transform shape}]
    \node (q0) {$\sim_{hhp}$};
    \node[right=7mm of q0] (q1) {$\sim_{hp}$};
    \node[right=7mm of q1] (q2) {$\sim_{p}$};
    \node[right=7mm of q2] (q3) {$\lesssim_{rp}$};
    \node[right=7mm of q3] (q4) {$\lesssim_{rt}$};
    \node[above right=7mm of q4] (q5) {$\lesssim_{ft}$};
    \node[below right=7mm of q4] (q6) {$\lesssim_{r}$};
    \node[right=15mm of q4] (q7) {$\lesssim_{f}$};
    \node[right=7mm of q7] (q8) {$\lesssim_{ct}$};
    \node[right=7mm of q8] (q9) {$\lesssim_{at}$};
    \node[below=7mm of q0] (q10) {$\lesssim_{rhhp}$};
    \node[below=7mm of q10] (q11) {$\lesssim_{hhp}$};
    \node[below=7mm of q1] (q12) {$\lesssim_{rhp}$};
    \node[below=7mm of q12] (q13) {$\lesssim_{hp}$};
    \node[below=7mm of q3] (q14) {$\lesssim_{p}$};
    \draw (q0) edge[->] (q1);
    \draw (q1) edge[->] (q2);
    \draw (q2) edge[->] (q3);
    \draw (q3) edge[->] (q4);
    \draw (q4) edge[->] (q5);
    \draw (q4) edge[->] (q6);
    \draw (q5) edge[->] (q7);
    \draw (q6) edge[->] (q7);
    \draw (q7) edge[->] (q8);
    \draw (q8) edge[->] (q9);
    \draw (q0) edge[->] (q10);
    \draw (q10) edge[->] (q11);
    \draw (q1) edge[->] (q12);
    \draw (q12) edge[->] (q13);
    \draw (q3) edge[->] (q14);
  \end{tikzpicture}
  \caption{Inclusions of preorders in $\Theta$.}
  \label{IOP}
\end{figure}
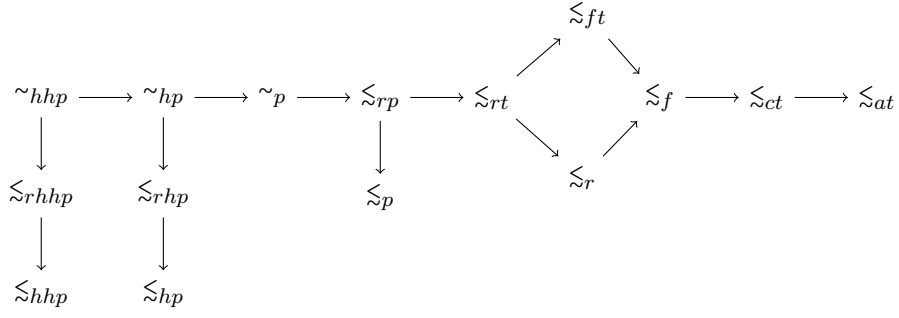

Moreover,

\begin{itemize}
  \item $\lesssim_{rhhp}\rightarrow\lesssim_{rhp}\rightarrow\lesssim_{rp}$.
  \item $\lesssim_{hhp}\rightarrow\lesssim_{hp}\rightarrow\lesssim_{p}$.
  \item $\lesssim_{p}\rightarrow\lesssim_{at}$ and $\lesssim_{p}\rightarrow\lesssim_{t}$.
  \item $\lesssim_{rp}\rightarrow\lesssim_{t}$.
  \item $\sim_{s}$ is a special case of $\sim_{p}$, $\lesssim_{s}$ is a special case of $\lesssim_{p}$ and $\lesssim_{rs}$ is a special case of $\lesssim_{rp}$.
\end{itemize}

The same inclusions hold for the kernels of the preorders.
\end{proposition}

\subsection{Baldan-Crafa Logic}\label{bcl}

Hennessy-Milner logic (HML) links logic of reactive programs and behavioural relations. Correspondingly, Baldan and Crafa \cite{ALTC} introduce multi-modal logics which characterize truly concurrent bisimulation equivalences. We call this truly concurrent version of Hennessy-Milner logic as Baldan-Crafa logic (BCL).

Let $\mathcal{P}$ be a PLTS and $\mathsf{Var}$ be a denumerable set of variables ranged over by $x,y,z$. We write $Env_{\mathcal{P}}$ as the set of $\eta:\mathsf{Var}\rightarrow\mathsf{Act}$. And we write $\mathbf{x}$ as the tuples of variables $x_1,\cdots,x_n\in \mathsf{Var}$ for $n\in\mathbb{N}$.

\begin{definition}[Syntax of Baldan-Crafa logic]
Given a PLTS $\mathcal{P}$, the set of BCL formulae is given by the BNF grammar:

$$\varphi::=\mathbf{true}|P|\neg\varphi|\varphi_1\wedge\varphi_2|(\mathbf{x},\overline{\mathbf{y}}<a~z)\varphi|\langle z\rangle \varphi$$

where $a\in\mathsf{Act}$, $P\in\mathsf{Pred}$ and $\mathbf{x},\mathbf{y}$ are tuples of variables, $z\in \mathsf{Var}$.
\end{definition}

\begin{definition}[Semantics of Baldan-Crafa logic]\label{SemanticsOfBCL}
Given a PLTS $\mathcal{P}$, let $\mathbf{C},\mathbf{C}'\in\mathcal{C}(\mathcal{P})$. $\mathcal{P}$ and $\mathbf{C}$ satisfies BCL formulae $\varphi$ in some environment $\eta$, denoted $(\mathcal{P},\mathbf{C})\models_{\eta}\varphi$, is defined inductively as follows:

$(\mathcal{P},\mathbf{C})\models_{\eta}\mathbf{true}$

$(\mathcal{P},\mathbf{C})\models_{\eta} P\Leftrightarrow \mathbf{C}P$

$(\mathcal{P},\mathbf{C})\models_{\eta}\neg\varphi\Leftrightarrow \textrm{ not }(\mathcal{P},\mathbf{C})\models_{\eta}\varphi,\varphi\textrm{ is closed}$

$(\mathcal{P},\mathbf{C})\models_{\eta}\varphi_1\wedge\varphi_2\Leftrightarrow (\mathcal{P},\mathbf{C})\models_{\eta}\varphi_1\textrm{ and }(\mathcal{P},\mathbf{C})\models_{\eta}\varphi_2$

$(\mathcal{P},\mathbf{C})\models_{\eta}(\mathbf{x},\overline{\mathbf{y}}<a~z)\varphi\Leftrightarrow \exists a\in\mathsf{Act}\setminus\mathbf{C}, \eta(\mathbf{x})<a,\eta(\mathbf{y})\parallel a\textrm{ and }(\mathcal{P},\mathbf{C})\models_{\eta[z\mapsto a]}\varphi$

$(\mathcal{P},\mathbf{C})\models_{\eta}\langle z\rangle\varphi\Leftrightarrow (\mathcal{P},\mathbf{C}')\models_{\eta}\varphi\textrm{ for some }\mathbf{C}'\textrm{ such that }\mathbf{C}\xrightarrow{\eta(z)}\mathbf{C}'$
\end{definition}

Two configurations $\mathbf{C},\mathbf{C}'$ are equivalent w.r.t. BCL, denoted $\mathbf{C}\sim_{BCL}\mathbf{C}'$ if and only if for all BCL formulae $\varphi$, $(\mathcal{P},\mathbf{C})\models_{\eta}\varphi\Leftrightarrow(\mathcal{P},\mathbf{C}')\models_{\eta}\varphi$. Then the following theorem holds.

\begin{theorem}
$\sim_{hhp}$ and $\sim_{BCL}$ coincide over finitely branching PLTSs.
\end{theorem}

We will write $\semangle{\mathbf{x},\overline{\mathbf{y}}< a~z}\varphi$ for the formulae $(\mathbf{x},\overline{\mathbf{y}}<a~z)\langle z\rangle\varphi$, $((\mathbf{x},\overline{\mathbf{y}}<a~z)\otimes(\mathbf{x}',\overline{\mathbf{y}'}<b~z'))\varphi$ for the formulae $(\mathbf{x},\overline{\mathbf{y}}<a~z)(\mathbf{x}',\overline{\mathbf{y}',z}<b~z')\varphi$, $(\semangle{\mathbf{x},\overline{\mathbf{y}}<a~z}\otimes\semangle{\mathbf{x}',\overline{\mathbf{y}'}<b~z'})\varphi$ for the formulae $(\mathbf{x},\overline{\mathbf{y}}<a~z)(\mathbf{x}',\overline{\mathbf{y}',z}<b~z')\langle z\rangle\langle z'\rangle\varphi$.

Then, we can define fragments of BCL for step logic (SL), pomset logic (PL) and hp logic (HPL) as follows.

Step logic: $\varphi::=\mathbf{true}|P|\neg\varphi|\varphi_1\wedge\varphi_2|(\semangle{a_1~x_1}\otimes\cdots\otimes\semangle{a_n~x_n})\varphi$

Pomset logic: $\varphi::=\mathbf{true}|P|\neg\varphi|\varphi_1\wedge\varphi_2|\semangle{\mathbf{x},\overline{\mathbf{y}}<a~z}\varphi$, where $\neg,\wedge$ are used only on closed formulae.

Hp logic: $\varphi::=\mathbf{true}|P|\neg\varphi|\varphi_1\wedge\varphi_2|\semangle{\mathbf{x},\overline{\mathbf{y}}<a~z}\varphi$

The semantics of step logic, pomset logic and hp logic are also defined in \cref{SemanticsOfBCL}.

Two configurations $\mathbf{C},\mathbf{C}'$ are equivalent w.r.t. step logic, denoted $\mathbf{C}\sim_{SL}\mathbf{C}'$ if and only if for all SL formulae $\varphi$, $(\mathcal{P},\mathbf{C})\models_{\eta}\varphi\Leftrightarrow(\mathcal{P},\mathbf{C}')\models_{\eta}\varphi$. Then the following theorem holds.

\begin{theorem}
$\sim_{s}$ and $\sim_{SL}$ coincide over finitely branching PLTSs.
\end{theorem}

Two configurations $\mathbf{C},\mathbf{C}'$ are equivalent w.r.t. pomset logic, denoted $\mathbf{C}\sim_{PL}\mathbf{C}'$ if and only if for all PL formulae $\varphi$, $(\mathcal{P},\mathbf{C})\models_{\eta}\varphi\Leftrightarrow(\mathcal{P},\mathbf{C}')\models_{\eta}\varphi$. Then the following theorem holds.

\begin{theorem}
$\sim_{p}$ and $\sim_{PL}$ coincide over finitely branching PLTSs.
\end{theorem}

Two configurations $\mathbf{C},\mathbf{C}'$ are equivalent w.r.t. hp logic, denoted $\mathbf{C}\sim_{HPL}\mathbf{C}'$ if and only if for all HPL formulae $\varphi$, $(\mathcal{P},\mathbf{C})\models_{\eta}\varphi\Leftrightarrow(\mathcal{P},\mathbf{C}')\models_{\eta}\varphi$. Then the following theorem holds.

\begin{theorem}
$\sim_{hp}$ and $\sim_{HPL}$ coincide over finitely branching PLTSs.
\end{theorem}

\subsection{Term Algebras}\label{ta}

Assume a countably infinite set $\mathsf{Var}$ of variables, ranged over by $x,y,z$.

\begin{definition}[Signature]
A signature $\Sigma$ is a set of function symbols, disjoint from $\mathsf{Var}$, together with an arity mapping that assigns a natural number $ar(f)$ to each function symbol $f$. Function symbols of arity zero, one and two are called constant, unary and binary, respectively.
\end{definition}

\begin{definition}[Term]
The set $\mathbb{T}(\Sigma)$ of (open) terms over a signature $\Sigma$, ranged over by $t,u$, is the least set such that:

\begin{itemize}
  \item Each $x\in \mathsf{Var}$ is a term.
  \item If $f$ is a function symbol and $t_1,\cdots,t_{ar(f)}$ are terms, then $f(t_1,\cdots,t_{ar(f)})$ is a term.
\end{itemize}

A term that does not contain any variable is called a closed term, the set of closed terms over a signature $\Sigma$ is denoted $T(\Sigma)$.
\end{definition}

\begin{definition}[Substitution]
A substitution is a mapping $\sigma:\mathsf{Var}\rightarrow\mathbb{T}(\Sigma)$ over $\Sigma$. $\sigma$ is closed if it maps each variable to a closed term in $T(\Sigma)$. $\sigma$ can be extended to $\sigma:\mathbb{T}(\Sigma)\rightarrow\mathbb{T}(\Sigma)$: $\sigma(t)$ is obtained by replacing occurrences of variables $x$ in $t$ by $\sigma(x)$.
\end{definition}

\begin{definition}[Context]
A context $C[x_1,\cdots,x_n]$ denotes an open term in which at most the distinct variables $x_1,\cdots,x_n$ appear. The term $C[t_1,\cdots,t_n]$ is obtained by replacing all occurrences of variables $x_i$ in $C[x_1,\cdots,x_n]$ by $t_i$ for $i=1,\cdots,n$.
\end{definition}

\begin{definition}[Congruence]
Assume a signature $\Sigma$. An equivalence relation (resp. preorder) $\mathcal{R}$ over $T(\Sigma)$ is a congruence (resp. precongruence), if for all $f\in\Sigma$,

$$t_i\mathcal{R}u_i\textrm{ for }i=1,\cdots,ar(f)\textrm{ implies }f(t_1,\cdots,t_{ar(f)})\mathcal{R}f(u_1,\cdots,u_{ar(f)})$$
\end{definition}

\subsection{Pomset Transition System Specifications}\label{ptss}

For a PLTS, the set $\mathsf{Proc}$ of states can be represented by not only the closed terms over some signature (the future of current state), but also the configuration set (the past of current state). A pomset transition system specification is a collection of inductive proof rules to derive the pomset transitions over states in $\mathsf{Proc}$.

\begin{definition}[Pomset transition system specification]
Let $\Sigma$ be a signature, and $t,t'\in\mathbb{T}(\Sigma)$. A pomset transition rule $\rho$ is of the form $\frac{H}{\alpha}$, where $H$ is the set of premises with positive premises $t\xrightarrow{U}t'$ and $tP$, and negative premises $t\xnrightarrow{U}$ and $t\neg P$; $\alpha$ is the conclusion with the form $t\xrightarrow{U}t'$ and $tP$, and $U\in\mathsf{Pom}$ and $P$ is a predicate. For the conclusion with the form of $t\xrightarrow{U}t'$, the left-hand side of the conclusion $t$ is called the source of $\rho$ and the right-hand side of the conclusion $t'$ is called the target of $\rho$. A transition rule is closed if it does not contain variables.

A pomset transition system specification (PTSS) is a set of pomset transition rules. A PTSS is positive if its transition rules do not contain negative premises.
\end{definition}

A literal is a positive literal with the pomset transition $t\xrightarrow{U}t'$ and $tP$, or a negative literal with the form $t\xnrightarrow{U}$ and $t\neg P$, where $U\in\mathsf{Pom}$, $P$ is a predicate, and $t,t'$ range over the collection of closed terms for some signature $\Sigma$.

\begin{definition}[Proof]\label{proof}
Let $T$ be a PTSS. A proof of a closed pomset transition rule $\frac{H}{\alpha}$ from $T$ is a well-founded, upwardly branching tree whose nodes are labelled by literals, its root is labelled by $\alpha$, and if $K$ is the set of labels of the nodes directly above a node with label $\beta$, then 

\begin{enumerate}
  \item Either $K=\emptyset$ and $\beta\in H$.
  \item Or $\frac{K}{\beta}$ is a closed substitution instance of a transition rule in $T$.
\end{enumerate}

If a proof of $\frac{H}{\alpha}$ from $T$ exists, then $\frac{H}{\alpha}$ is provable from $T$, denoted $T\vdash\frac{H}{\alpha}$.
\end{definition} 

\subsection{Examples of PTSSs}\label{eptss}

\subsubsection{Basic Process Algebra with Empty Process}\label{bpawep}

The signature of Basic Process Algebra with Empty Process, denoted by $BPA_{\epsilon}$, consists of the following operators: 

\begin{itemize}
  \item A set of constants, denoted by $\mathsf{Act}$, and a set of pomset over $\mathsf{Act}$, denoted by $\mathsf{Pom}$.
  \item A special empty process constant, denoted by $\epsilon$. 
  \item A binary alternative composition, denoted by $+$.
  \item A binary sequential composition, denoted by $\cdot$.
\end{itemize} 

The BNF grammar for $BPA_{\epsilon}$ is as follows.

$$t::=U|\epsilon|t_1+t_2|t_1\cdot t_2$$

Where $U\in\mathsf{Pom}$. The PTSS of $BPA_{\epsilon}$ is shown in \cref{TRForBPAEpsilon}, where $U\in\mathsf{Pom}$, $\surd$ is a predicate representing successful termination, and $x,x',y,y'\in T(BPA_{\epsilon})$.

\begin{center}
    \begin{table}
        $$\frac{}{U\xrightarrow{U}\epsilon}\quad\frac{}{\epsilon\surd}$$
        $$\frac{x\surd}{x+ y\surd} \quad\frac{x\xrightarrow{U}x'}{x+ y\xrightarrow{U}x'} \quad\frac{y\surd}{x+ y\surd} \quad\frac{y\xrightarrow{U}y'}{x+ y\xrightarrow{U}y'}$$
        $$\frac{x\surd\quad y\surd}{x\cdot y\surd}\quad\frac{x\surd\quad y\xrightarrow{U}y'}{x\cdot y\xrightarrow{U} y'} \quad\frac{x\xrightarrow{U}x'}{x\cdot y\xrightarrow{U}x'\cdot y}$$
        \caption{Pomset transition rules of $BPA_{\epsilon}$}
        \label{TRForBPAEpsilon}
    \end{table}
\end{center}

\subsubsection{Parallelism}\label{parallelism}

The signature Algebra of Parallelism for True Concurrency, denoted by $APTC$, is obtained by adding the parallel operator $\parallel$ from \cite{APTC} to $BPA_{\epsilon}$.

The BNF grammar for $APTC$ is as follows.

$$t::=U|\epsilon|t_1+t_2|t_1\cdot t_2|t_1\parallel t_2$$

Where $U\in\mathsf{Pom}$. The PTSS of $APTC$ includes pomset transition rules of $BPA_{\epsilon}$ in \cref{TRForBPAEpsilon} and pomset transition rules of $\parallel$ in \cref{TRForAPTC}, where $U,U_1,U_2\in\mathsf{Pom}$, $\surd$ is a predicate representing successful termination, and $x,x',y,y'\in T(APTC)$.

\begin{center}
    \begin{table}
        $$\frac{x\surd\quad y\surd}{x\parallel y\surd} \quad\frac{x\xrightarrow{U}x'\quad y\surd}{x\parallel y\xrightarrow{U}x'}$$
        $$\frac{x\surd\quad y\xrightarrow{U}y'}{x\parallel y\xrightarrow{U}y'} \quad\frac{x\xrightarrow{U_1}x'\quad y\xrightarrow{U_2}y'}{x\parallel y\xrightarrow{U_1\parallel U_2}x'\parallel y'}$$
        \caption{Pomset transition rules of $APTC$}
        \label{TRForAPTC}
    \end{table}
\end{center} 

\subsubsection{Priority}\label{priority}

The signature of Basic Process Algebra with Empty Process and Priority, denoted by $BPA_{\epsilon\theta}$, is obtained by adding the unary priority operator $\theta$ to $BPA_{\epsilon}$. $\theta$ defines a priority partial order on $\mathsf{Pom}$, denoted by $<_p$, distinct from the partial orders in a pomset $U\in\mathsf{Pom}$.

The BNF grammar for $BPA_{\epsilon\theta}$ is as follows.

$$t::=U|\epsilon|t_1+t_2|t_1\cdot t_2|\theta(t)$$

Where $U\in\mathsf{Pom}$. The PTSS of $BPA_{\epsilon\theta}$ includes pomset transition rules of $BPA_{\epsilon}$ in \cref{TRForBPAEpsilon} and pomset transition rules of $\theta$ in \cref{TRForBPAEpsilonPriority}, where $U,U_1,U_2\in\mathsf{Pom}$, $\surd$ is a predicate representing successful termination, and $x,x',y,y'\in T(BPA_{\epsilon\theta})$.

\begin{center}
    \begin{table}
        $$\frac{x\surd}{\theta(x)\surd} \quad\frac{x\xrightarrow{U_1}x'\quad x\xnrightarrow{U_2}\quad U_1<_p U_2}{\theta(x)\xrightarrow{U_1}\theta(x')}$$
        \caption{Pomset transition rules of $\theta$}
        \label{TRForBPAEpsilonPriority}
    \end{table}
\end{center} 

\subsubsection{Discrete Time}\label{discretetime}

The signature of Basic Process Algebra with Empty Process and Discrete Time, denoted by $BPA_{\epsilon}^{dt}$, is obtained by adding the unary operator $\sigma_d$ (representing a delay of one time slice) to $BPA_{\epsilon}$. 

The BNF grammar for $BPA_{\epsilon}^{dt}$ is as follows.

$$t::=U|\epsilon|t_1+t_2|t_1\cdot t_2|\sigma_d(t)$$

Where $U\in\mathsf{Pom}$. The PTSS of $BPA_{\epsilon}^{dt}$ includes pomset transition rules of $BPA_{\epsilon}$ in \cref{TRForBPAEpsilon} and pomset transition rules of $\sigma_d$ in \cref{TRForBPAEpsilonDT}, where $U,U_1,U_2\in\mathsf{Pom}$, $\surd$ is a predicate representing successful termination, and $x,x',y,y'\in T(BPA_{\epsilon}^{dt})$.

\begin{center}
    \begin{table}
        $$\frac{}{\sigma_d(x)\xrightarrow{\sigma}x}$$
        $$\frac{x\xrightarrow{\sigma}x'\quad y\xnrightarrow{\sigma}}{x+ y\xrightarrow{\sigma}x'} \quad\frac{y\xrightarrow{\sigma}y'\quad x\xnrightarrow{\sigma}}{x+ y\xrightarrow{\sigma}y'}
        \quad \frac{x\xrightarrow{\sigma}x'\quad y\xrightarrow{\sigma}y'}{x+ y\xrightarrow{\sigma}x'+y'}$$
        $$\frac{x\surd\quad y\xrightarrow{\sigma}y'}{x\cdot y\xrightarrow{\sigma} y'} \quad\frac{x\xrightarrow{\sigma}x'}{x\cdot y\xrightarrow{\sigma}x'\cdot y}$$
        \caption{Pomset transition rules of $\sigma_d$}
        \label{TRForBPAEpsilonDT}
    \end{table}
\end{center} 
\newpage\section{The Meaning of PLTSs}\label{tmop}

Firstly, let us consider the negative pomset transitions. The pomset transition $t\neg P$ is same to traditional single event one. While $t\xnrightarrow{U}t'$ and $t\xnrightarrow{U}$ for $U\in\mathsf{Pom}$ are quite different to the traditional single event ones. From \cref{LemmaFactorization}, we know that a pomset $U\in\mathsf{Pom}$ can be factorized into a sequential factorization or a parallel factorization. And from the so-called exchange law modulo truly concurrent bisimilarities \cite{APTC} ($(a\cdot t)\parallel(b\cdot u)=(a\parallel b)\cdot(t\parallel u)$ for $a,b\in\mathsf{Act}$ and $t,u\in\mathbb{T}(\Sigma)$ for some signature $\Sigma$), a pomset $U\in\mathsf{Pom}$ can be factorized into a sequence of steps $U=U_1\cdot\cdots\cdot U_i\cdot\cdots\cdot U_n$ with each $U_i=a_{i_1}\parallel\cdots\parallel a_{i_m}$, where $m,n\in\mathbb{N}$, $i=1,\cdots,n$ and $a_{i_1},\cdots,a_{i_m}\in\mathsf{Act}$. Generally, $t\xnrightarrow{U}t'$ has the form $t\xrightarrow{U_1}t_1\xrightarrow{U_2}\cdots\xrightarrow{U_{i-1}} t_{i-1}\xnrightarrow{U_i}t_i\xnrightarrow{U_{i+1}}\cdots\xnrightarrow{U_n}t'$ while $t\xnrightarrow{U}$ has the form $t\xrightarrow{U_1}\xrightarrow{U_2}\cdots\xrightarrow{U_{i-1}}\xnrightarrow{U_i}\xnrightarrow{U_{i+1}}\cdots\xnrightarrow{U_n}$. For simplicity, we say $t\xnrightarrow{U}t'$ and $t\xnrightarrow{U}$, we mean $t$ can not execute the first step $U_1$ of $U$, i.e., $t\xnrightarrow{U_1}t'$ and $t\xnrightarrow{U_1}$. 

Then, about the two questions: (1) which PTSSs are meaningful? (2) which PLTSs can be associated with them? The answers are almost same to those of TSSs and LTSs, we do not follow the whole history like in \cite{SOS}, and only write some main definitions and conclusions.

Similarly to \cite{SOS}, we answer the above two questions in three aspects: in \cref{mta}, we give the model-theoretic answers; in \cref{pta}, we give the proof-theoretic answers; we give the answers based on stratification in \cref{abos}.

\subsection{Model-Theoretic Answers}\label{mta}

The concept of three-valued stable model is important to the congruence theorem in the presence of negative premises. A three-valued stable model partitions the collection of transitions into three disjoint sets: the set $C$ of transitions that are true, the set $F$ of transitions that are false and the set $V$ of transitions for which it is unknown whether or not they are true. Such a partitioning is determined by the pair $(C,V)$.

\begin{definition}
A set $N$ of negative pomset transitions $t\xnrightarrow{U}$ and $t\neg P$ holds for a set $\mathcal{S}$ of transitions, denoted by $\mathcal{S}\models N$, if the following conditions hold:

\begin{enumerate}
  \item For each $t\xnrightarrow{U}\in N$, $t\xrightarrow{U}t'\notin \mathcal{S}$.
  \item For each $t\neg P\in N$, $tP\notin\mathcal{S}$.
\end{enumerate}

Where $t,t'\in T(\Sigma)$ for some signature $\Sigma$, $U\in\mathsf{Pom}$ and $P\in\mathsf{Pred}$.
\end{definition}

\begin{definition}[Three-valued stable model]
A pair $(C,V)$ of disjoint sets of transitions is a three-valued stable model for a PTSS $T$, if the following conditions hold:

\begin{enumerate}
  \item A transition $\alpha\in C$ if and only if $T$ proves a closed transition rule $\frac{N}{\alpha}$ where $N$ contains only negative premises and $C\cup V\models N$.
  \item A transition $\alpha\in C\cup V$ if and only if $T$ proves a closed transition rule $\frac{N}{\alpha}$ where $N$ contains only negative premises and $C\models N$.
\end{enumerate}
\end{definition}

Each PTSS $T$ allows a least three-valued stable model $(C,V)$ in the sense that the sets $C$ and $F$ are both minimal and the set $V$ is maximal. A PTSS is meaningful if and only if its least three-valued stable model does not contain any unknown transitions and is called positive after reduction.

\subsection{Proof-Theoretic Answers}\label{pta}

We introduce the answers based on the concept of proof. The concept of well-supported (ws) model has a deep relation to three-valued stable model with negative premises.

\begin{definition}[Denying literals]
Denying literals is a pair of literals that deny each other with $U\in\mathsf{Pom}$ and $P\in\mathsf{Pred}$:

\begin{itemize}
  \item $t\xrightarrow{U}t'$ and $t\xnrightarrow{U}t'$.
  \item $t\xrightarrow{U}t'$ and $t\xnrightarrow{U}$.
  \item $tP$ and $t\neg P$.
\end{itemize}
\end{definition}

\begin{definition}[Well-supported proof]
A well-supported proof of a literal $\alpha$ from a PTSS $T$ is like a proof defined in \cref{proof} with an extra condition:

\begin{enumerate}
  \setcounter{enumi}{2}
  \item $\beta$ is negative, and for each set $N$ of negative literals such that $T\vdash \frac{N}{\gamma}$, where $\gamma$ is a literal denying $\beta$, a literal in $N$ denies one in $K$.
\end{enumerate}

If there is a well-supported proof of $\alpha$ from $T$, we write $T\vdash_{ws}\alpha$.
\end{definition}

\begin{proposition}
For any PTSS $T$ and literal $\alpha$, the induced relation $\vdash_{ws}$ does not contain denying literals, and $T\vdash_{ws}\alpha$ implies $L\models\alpha$ for each well-supported model $L$ of $T$.
\end{proposition}

\begin{definition}[Ws-completeness]
A PTSS $T$ is ws-complete if for any pomset transition $t\xrightarrow{U}t'$ (resp. $tP$) either $T\vdash_{ws}t\xrightarrow{U}t'$ (resp. $T\vdash_{ws}tP$) or $T\vdash_{ws}t\xnrightarrow{U}t'$ (resp. $T\vdash_{ws}t\neg P$).
\end{definition}

\begin{proposition}
A PTSS is ws-complete if and only if its least three-valued stable model does not contain any unknown transitions, i.e., it is positive after reduction.
\end{proposition}

\subsection{Answers Based on Stratification}\label{abos}

\begin{definition}[Ordinal number]
The ordinal numbers are defined inductively as follows:

\begin{enumerate}
  \item $0$ is the smallest ordinal number.
  \item Each ordinal number $\xi$ has a successor $\xi+1$.
  \item Each sequence of ordinal numbers $\xi<\xi+1<\xi+2<\cdots$ is capped by a limit ordinal $\lambda$.
\end{enumerate}
\end{definition}

\begin{definition}[Stratification]
A stratification for a PTSS is a weight function $\phi$ which maps transitions to ordinal numbers, such that for each pomset transition rule $\rho$ with conclusion $\alpha$ and for each closed substitution $\sigma$:

\begin{enumerate}
  \item For positive premises $t\xrightarrow{U}t'$ and $tP$ of $\rho$, $\phi(\sigma(t)\xrightarrow{U}\sigma(t'))\leq\phi(\sigma(\alpha))$ and $\phi(\sigma(t)P)\leq\phi(\sigma(\alpha))$, respectively.
  \item For negative premises $t\xnrightarrow{U}$ and $t\neg P$ of $\rho$, $\phi(\sigma(t)\xrightarrow{U}t')<\phi(\sigma(\alpha))$ and $\phi(\sigma(t)P)<\phi(\sigma(\alpha))$, respectively.
\end{enumerate}
\end{definition}

\begin{proposition}
If a PTSS allows a stratification, then its least three-valued stable model does not contain any unknown transitions, i.e., it is positive after reduction.
\end{proposition} 

Similar conclusions to those in \cite{SOS} can be drawn as follows when the above concepts applied to the concrete algebras in \cref{eptss}:

\begin{enumerate}
  \item The PTSS of $BPA_{\epsilon}$ in \cref{bpawep} is positive.
  \item The PTSS of $APTC$ in \cref{parallelism} is positive.
  \item The PTSS of $BPA_{\epsilon\theta}$ in \cref{priority} is positive after reduction and ws-complete, because it allows a suitable stratification.
  \item The PTSS of $BPA^{dt}_{\epsilon}$ in \cref{discretetime} is positive after reduction and ws-complete, because it allows a suitable stratification.
\end{enumerate}
\newpage\section{Conservative Extensions}\label{ce} 

An operational conservative extension requires that an original PTSS and its extension prove exactly the same closed transition rules that have only negative premises and an original closed term as their source. In \cref{oce}, we introduce the concept and main conclusion of operational conservative extension. In \cref{iftvsm}, we introduce the implications of operational conservative extension for three-valued stable models. Then, we introduce applications of conservative extensions to axiomatizations and term rewrite systems in \cref{aoa} and \cref{aor}, respectively.

\subsection{Operational Conservative Extension}\label{oce}

\begin{definition}[Operational conservative extension]
$\Sigma_0$ and $\Sigma_1$ are two signatures, the union of $\Sigma_0$ and $\Sigma_1$, denoted $\Sigma_0\oplus\Sigma_1$, agrees on the arity of the function symbols in the intersection of $\Sigma_0$ and $\Sigma_1$. $T_0$ and $T_1$ are PTSSs over $\Sigma_0$ and $\Sigma_1$, the sum of $T_0$ and $T_1$, denoted $T_0\oplus T_1$, is the PTSS over signature $\Sigma_0\oplus\Sigma_1$ containing the pomset transition rules in $T_0$ and $T_1$. A PTSS $T_0\oplus T_1$ is an operational conservative extension of PTSS $T_0$, if for each closed transition rule $\frac{N}{\alpha}$ such that:

\begin{itemize}
  \item $N$ contains only negative literals.
  \item The left-hand side of $\alpha$ is in $T(\Sigma_0)$.
  \item $T_0\oplus T_1\vdash\frac{N}{\alpha}$.
\end{itemize}

we have that $T_0\vdash\frac{N}{\alpha}$.
\end{definition}

\begin{definition}[Source-dependency]
The source-dependent variables in a pomset transition rule $\rho$ are defined inductively as follows:

\begin{itemize}
  \item All variables in the source of $\rho$ are source-dependent.
  \item If $t\xrightarrow{U}t'$ is a premise of $\rho$ and all variables in $t$ are source-dependent, then all variables in $t'$ are source-dependent.
\end{itemize}

A pomset transition rule is source-dependent if all its variables are.
\end{definition}

\begin{definition}[Freshness]
We say that a term in $\mathbb{T}(\Sigma_0\oplus\Sigma_1)$ is fresh if it contains a function symbol (may be a constant) from $\Sigma_1\setminus\Sigma_0$. Similarly, a predicate symbol in $T_1$ is fresh if it does not occur in $T_0$.
\end{definition}

\begin{theorem}[Operational conservative extension]
Let $T_0$ and $T_1$ be PTSSs over signature $\Sigma_0$ and $\Sigma_1$, respectively. $T_0\oplus T_1$ is an operational conservative extension of $T_0$, if:

\begin{enumerate}
  \item Each $\rho\in T_0$ is source-dependent.
  \item For each $\rho\in T_1$,
       \begin{itemize}
         \item Either the source of $\rho$ is fresh.
         \item Or $\rho$ has a premise of the form $t\xrightarrow{U}t'$ or $tP$, where:
              \begin{itemize}
                \item $t\in\mathbb{T}(\Sigma_0)$.
                \item All variables in $t$ occur in the source of $\rho$.
                \item $t'$, $P$ and every $a\in U$ are fresh.
              \end{itemize}
       \end{itemize}
\end{enumerate}
\end{theorem}

\subsection{Implications for Three-Valued Stable Models}\label{iftvsm}

\begin{proposition}
Let $T_0\oplus T_1$ be an operational conservative extension of $T_0$. If $(C,V)$ is a three-valued stable model of $T_0\oplus T_1$, then,

$$C'\triangleq \{\alpha\in C|\textrm{the left-hand side of }\alpha\textrm{ is in }T(\Sigma_0)\}$$
$$V'\triangleq \{\alpha\in V|\textrm{the left-hand side of }\alpha\textrm{ is in }T(\Sigma_0)\}$$

$(C',V')$ is a three-valued stable model of $T_0$.
\end{proposition}

\begin{proposition}
Let $T_0\oplus T_1$ be an operational conservative extension of $T_0$. If $(C,V)$ is a three-valued stable model of $T_0$, then there exists a three-valued stable model $(C',V')$ of $T_0\oplus T_1$ such that

$$C\triangleq \{\alpha\in C'|\textrm{the left-hand side of }\alpha\textrm{ is in }T(\Sigma_0)\}$$
$$V\triangleq \{\alpha\in V'|\textrm{the left-hand side of }\alpha\textrm{ is in }T(\Sigma_0)\}$$
\end{proposition}

\begin{corollary}
Let $T_0\oplus T_1$ be an operational conservative extension of $T_0$. If $(C,V)$ is the least three-valued stable model of $T_0\oplus T_1$, then,

$$C'\triangleq \{\alpha\in C|\textrm{the left-hand side of }\alpha\textrm{ is in }T(\Sigma_0)\}$$
$$V'\triangleq \{\alpha\in V|\textrm{the left-hand side of }\alpha\textrm{ is in }T(\Sigma_0)\}$$

$(C',V')$ is the least three-valued stable model of $T_0$.
\end{corollary}

\subsection{Applications to Axiomatizations}\label{aoa}

\subsubsection{Axiomatic Conservative Extension}

\begin{definition}[Axiomatization]
An axiomatization $\mathcal{E}$ over a signature $\Sigma$ consists of a set of equations, called axioms, of the form $t=u$ with $t,u\in\mathbb{T}(\Sigma)$. An axiomatization $\mathcal{E}$ induces a binary equality relation $=$ on $\mathbb{T}(\Sigma)$:

\begin{itemize}
  \item (Substitution) If $t=u$ is an axiom and $\sigma$ is a substitution, then $\sigma(t)=\sigma(u)$ is an axiom.
  \item (Equivalence) The relation $=$ is an equivalence, i.e., closed under reflexivity, symmetry and transitivity.
  \item (Context) The relation $=$ is closed under context, i.e., if $t=u$ is an axiom and $f$ is a function symbol with arity $ar(f)>0$, then,
  
  $$f(s_1,\cdots,s_{i-1},t,s_{i+1},\cdots,s_{ar(f)})=f(s_1,\cdots,s_{i-1},u,s_{i+1},\cdots,s_{ar(f)})$$
\end{itemize}
\end{definition}

\begin{definition}[Soundness and completeness]
Assume an axiomatization $\mathcal{E}$ over a signature $\Sigma$ and an equivalence $\sim$ on $T(\Sigma)$:

\begin{enumerate}
  \item $\mathcal{E}$ is sound modulo $\sim$ if and only if $t=u$ implies $t\sim u$ for all $t,u\in T(\Sigma)$.
  \item $\mathcal{E}$ is complete modulo $\sim$ if and only if $t\sim u$ implies $t=u$ for all $t,u\in T(\Sigma)$.
\end{enumerate}
\end{definition}

\begin{definition}[Axiomatic conservative extension]
Let $\mathcal{E}_0$ and $\mathcal{E}_1$ be axiomatizations over $\Sigma_0$ and $\Sigma_0\oplus\Sigma_1$, respectively. Then $\mathcal{E}_0\cup\mathcal{E}_1$ is an axiomatic conservative extension of $\mathcal{E}_0$ if every $t=u$ with $t,u\in T(\Sigma_0)$ that can be derived from $\mathcal{E}_0\cup\mathcal{E}_1$ can also be derived from $\mathcal{E}_0$.
\end{definition}

\begin{theorem}
Let $\mathcal{E}_0$ and $\mathcal{E}_1$ be axiomatizations over $\Sigma_0$ and $\Sigma_0\oplus\Sigma_1$, respectively. And let $\sim$ be an equivalence relation on $T(\Sigma_0\oplus\Sigma_1)$. If:

\begin{enumerate}
  \item $\mathcal{E}_0\cup\mathcal{E}_1$ is sound over $T(\Sigma_0\oplus\Sigma_1)$ modulo $\sim$.
  \item $\mathcal{E}_0$ is complete over $T(\Sigma_0)$ modulo $\sim$.
\end{enumerate}

Then $\mathcal{E}_0\cup\mathcal{E}_1$ is an axiomatic conservative extension of $\mathcal{E}_0$.
\end{theorem}

\subsubsection{Completeness of Axiomatizations}

\begin{theorem}
Let $\mathcal{E}_0$ and $\mathcal{E}_1$ be axiomatizations over $\Sigma_0$ and $\Sigma_0\oplus\Sigma_1$, respectively. And let $\sim$ be an equivalence relation on $T(\Sigma_0\oplus\Sigma_1)$. If $\mathcal{E}_0\cup\mathcal{E}_1$ is an axiomatic conservative extension of $\mathcal{E}_0$, and for each $t\in T(\Sigma_0\oplus\Sigma_1)$ there is a $t'\in T(\Sigma_0)$ such that $t=t'$ can be derived from $\mathcal{E}_0\cup\mathcal{E}_1$, then $\mathcal{E}_0\cup\mathcal{E}_1$ is complete over $T(\Sigma_0\oplus\Sigma_1)$ modulo $\sim$.
\end{theorem}

\subsubsection{$\omega$-Completeness of Axiomatizations}

\begin{definition}[$\omega$-completeness]
An axiomatization $\mathcal{E}$ over a signature $\Sigma$ is $\omega$-complete if $t=u$ with $t,u\in\mathbb{T}(\Sigma)$ can be derived from $\mathcal{E}$ whenever $\sigma(t)=\sigma(u)$ can be derived from $\mathcal{E}$ for all closed substitution $\sigma$.
\end{definition}

\begin{theorem}
Let $\sim$ be an equivalence relation on $\mathbb{T}(\Sigma)$ for some signature $\Sigma$. Suppose that for all $t,u\in\mathbb{T}(\Sigma)$, $t\sim u$ whenever $\sigma(t)\sim\sigma(u)$ for all closed substitution $\sigma$. If $\mathcal{E}$ is an axiomatization over $\Sigma$ such that:

\begin{enumerate}
  \item $\mathcal{E}$ is sound over $T(\Sigma)$ modulo $\sim$.
  \item $\mathcal{E}$ is complete over $\mathbb{T}(\Sigma)$ modulo $\sim$.
\end{enumerate}

Then $\mathcal{E}$ is $\omega$-complete.
\end{theorem}

\subsection{Applications to Rewriting}\label{aor}

\subsubsection{Rewrite Conservative Extension}

\begin{definition}[Term rewrite system]
Assume a signature $\Sigma$. A term rewrite system (TRS) over $\Sigma$ consists of a set of rewrite rules $t\rightarrow u$ with $t,u\in\mathbb{T}(\Sigma)$. A TRS $\mathcal{R}$ induces a binary equality relation $\rightarrow$ on $\mathbb{T}(\Sigma)$:

\begin{itemize}
  \item (Substitution) If $t\rightarrow u$ is a rewrite rule and $\sigma$ is a substitution, then $\sigma(t)\rightarrow\sigma(u)$ is a rewrite rule.
  \item (Reflexivity and Transitivity) The relation $\rightarrow$ is closed under reflexivity and transitivity.
  \item (Context) The relation $\rightarrow$ is closed under context, i.e., if $t\rightarrow u$ is a rewrite rule and $f$ is a function symbol with arity $ar(f)>0$, then,
  
  $$f(s_1,\cdots,s_{i-1},t,s_{i+1},\cdots,s_{ar(f)})\rightarrow f(s_1,\cdots,s_{i-1},u,s_{i+1},\cdots,s_{ar(f)})$$
\end{itemize}
\end{definition}

$\rightarrow^*$ is the reflexive and transitive closure of $\rightarrow$.

\begin{definition}[Rewrite conservative extension]
Let $\mathcal{R}_0$ and $\mathcal{R}_1$ be TRSs over $\Sigma_0$ and $\Sigma_0\oplus\Sigma_1$, respectively. Then $\mathcal{R}_0\cup\mathcal{R}_1$ is a rewrite conservative extension of $\mathcal{R}_0$ if every $t\rightarrow^* u$ with $t,u\in T(\Sigma_0)$ that can be derived from $\mathcal{R}_0\cup\mathcal{R}_1$ can also be derived from $\mathcal{R}_0$.
\end{definition}

\begin{theorem}
Let $\mathcal{R}_0$ and $\mathcal{R}_1$ be TRSs over signature $\Sigma_0$ and $\Sigma_0\oplus\Sigma_1$, respectively. $\mathcal{R}_0\oplus \mathcal{R}_1$ is a rewrite conservative extension of $\mathcal{R}_0$, if:

\begin{enumerate}
  \item Each $\rho\in \mathcal{R}_0$ is source-dependent.
  \item For each $\rho\in \mathcal{R}_1$,
       \begin{itemize}
         \item Either the source of $\rho$ is fresh.
         \item Or $\rho$ has a premise of the form $t\rightarrow t'$, where:
              \begin{itemize}
                \item $t\in\mathbb{T}(\Sigma_0)$.
                \item All variables in $t$ occur in the source of $\rho$.
                \item $t'$ is fresh.
              \end{itemize}
       \end{itemize}
\end{enumerate}
\end{theorem}

\subsubsection{Ground Confluence of TRSs} 

\begin{definition}[Ground confluence]
A TRS is ground confluent if for all $t,t_0,t_1\in T(\Sigma)$ for some signature $\Sigma$ with $t\rightarrow^* t_0$ and $t\rightarrow^* t_1$, there is a $u\in T(\Sigma)$ with $t_0\rightarrow^* u$ and $t_1\rightarrow^* u$.
\end{definition}

A TRS $\mathcal{R}$ is sound modulo an equivalence relation $\sim$ on $T(\Sigma)$ if $t\rightarrow^* u$ implies $t\sim u$ for all $t,u\in T(\Sigma)$.

\begin{theorem}
Let $\sim$ be an equivalence relation on $T(\Sigma_0\oplus\Sigma_1)$ for some signatures $\Sigma_0$ and $\Sigma_1$. Assume TRSs $\mathcal{R}_0$ and $\mathcal{R}_1$ over $\Sigma_0$ and $\Sigma_0\oplus\Sigma_1$, respectively, such that:

\begin{enumerate}
  \item $\mathcal{R}_0\oplus\mathcal{R}_1$ is sound over $T(\Sigma_0\oplus\Sigma_1)$ modulo $\sim$.
  \item If $t,t'\in T(\Sigma_0)$ with $t\sim t'$, then there is a $u\in T(\Sigma_0)$ such that $t\rightarrow^* u$ and $t'\rightarrow^* u$ can be derived from $\mathcal{R}_0$.
  \item For each $t\in T(\Sigma_0\oplus\Sigma_1)$ there is a $t'\in T(\Sigma_0)$ such that $t\rightarrow^* t'$ can be derived from $\mathcal{R}_0\oplus\mathcal{R}_1$.
\end{enumerate}

Then $\mathcal{R}_0\oplus\mathcal{R}_1$ is ground confluent over $T(\Sigma_0\oplus\Sigma_1)$.
\end{theorem}

Similar conclusions to those in \cite{SOS} can be drawn as follows when the above concepts applied to the concrete algebras in \cref{eptss}:

\begin{enumerate}
  \item The PTSS of $BPA_{\epsilon}$ in \cref{bpawep} is source-dependent.
  \item $APTC$ in \cref{parallelism} is an operational conservative extension of $BPA_{\epsilon}$ in \cref{bpawep}.
  \item $BPA_{\epsilon\theta}$ in \cref{priority} is an operational conservative extension of $BPA_{\epsilon}$ in \cref{bpawep}.
  \item $BPA^{dt}_{\epsilon}$ in \cref{discretetime} is an operational conservative extension of $BPA_{\epsilon}$ in \cref{bpawep}.
\end{enumerate} 
\newpage\section{Congruence Formats}\label{cf} 

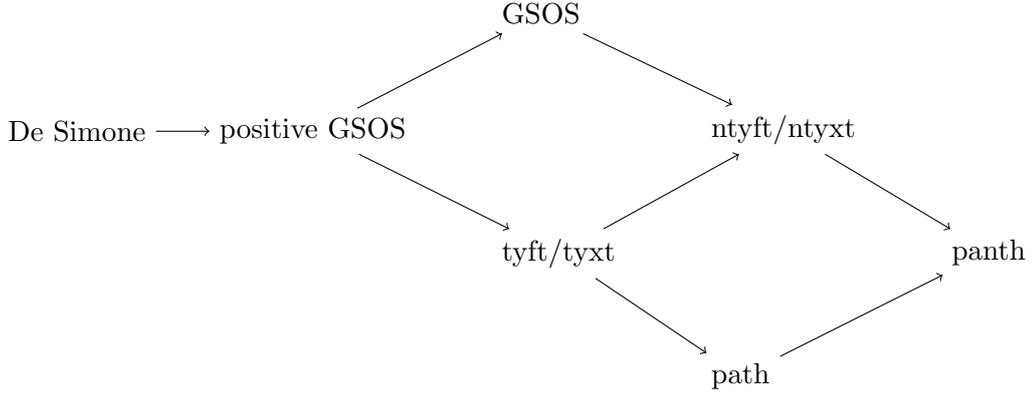
\begin{figure}
  \centering
  \begin{tikzpicture}[every node/.style={transform shape}]
    \node (q0) {De Simone};
    \node[right=7mm of q0] (q1) {positive GSOS};
    \node[above right=14mm of q1] (q2) {GSOS};
    \node[below right=14mm of q1] (q3) {tyft/tyxt};
    \node[above right=14mm of q3] (q4) {ntyft/ntyxt};
    \node[below right=14mm of q3] (q5) {path};
    \node[below right=14mm of q4] (q6) {panth};
    \draw (q0) edge[->] (q1);
    \draw (q1) edge[->] (q2);
    \draw (q1) edge[->] (q3);
    \draw (q2) edge[->] (q4);
    \draw (q3) edge[->] (q4);
    \draw (q3) edge[->] (q5);
    \draw (q4) edge[->] (q6);
    \draw (q5) edge[->] (q6);
  \end{tikzpicture}
  \caption{Lattice of rule formats.}
  \label{RF}
\end{figure}

A central issue of SOS is to define transition rule formats to ensure that a behavioural equivalence relation is a congruence. In the context of SOS for true concurrency, we will examine which traditional rule format ensures which truly concurrent behavioural equivalence being a congruence. The rule formats in \cref{RF} illustrate the relations among the well-known existed formats, while an arrow $\rightarrow$ from one format to another indicates that all transition rules in the first format are also in the second format and two formats without an arrow between them indicate that they are incomparable. 

We introduce panth format in \cref{panthformat}, ntree format in \cref{ntreeformat}, De Simone format in \cref{desimoneformat}, GSOS format in \cref{gsosformat}, RBB safe format in \cref{rbbformat}, respectively. Then, we introduce precongruence related formats in \cref{precongruence} and trace congruences in \cref{tracec}.

\subsection{Panth Format}\label{panthformat}

\begin{definition}[Panth format]
A pomset transition rule $\rho$ is in panth format if it satisfies the following three restrictions:

\begin{enumerate}
  \item For each positive premise $t\xrightarrow{U}t'$ of $\rho$ with $U\in\mathsf{Pom}$, the right-hand side $t'$ is a single variable.
  \item The source of $\rho$ contains no more than one function symbol.
  \item The variables that occur as right-hand sides of positive premises or in the source of $\rho$ are all distinct.
\end{enumerate}

A PTSS is in panth format if it consists of panth rules only.
\end{definition}

\begin{definition}[Path, ntyft/ntyxt and tyft/tyxt format]
A PTSS is in path format if it is in panth format and positive.

A PTSS is in ntyft/ntyxt format if it is in panth format and its pomset transition rules do not contain predicates.

A PTSS is in tyft/tyxt format if it is both path and ntyft/ntyxt.

A PTSS is in tyft format if it is tyft/tyxt and the source of each rule contains exactly one function symbol.
\end{definition}

\begin{theorem}
If a PTSS is ws-complete and panth, then pomset bisimulation equivalence is a congruence w.r.t. the PLTS associated with it.
\end{theorem}

\begin{theorem}
If a PTSS is ws-complete and panth, then step bisimulation equivalence is a congruence w.r.t. the PLTS associated with it.
\end{theorem}

\begin{theorem}
If a PTSS is ws-complete and panth, then hp-bisimulation equivalence is a congruence w.r.t. the PLTS associated with it.
\end{theorem}

\begin{theorem}
If a PTSS is ws-complete and panth, then hhp-bisimulation equivalence is a congruence w.r.t. the PLTS associated with it.
\end{theorem}

Similar conclusions to those in \cite{SOS} can be drawn as follows when the above concepts applied to the concrete algebras in \cref{eptss}:

\begin{enumerate}
  \item The PTSS of $BPA_{\epsilon}$ in \cref{bpawep} is panth.
  \item The PTSS of $APTC$ in \cref{parallelism} is panth.
  \item The PTSS of $BPA_{\epsilon\theta}$ in \cref{priority} is panth.
  \item The PTSS of $BPA^{dt}_{\epsilon}$ in \cref{discretetime} is panth.
\end{enumerate}

\subsection{Ntree Format}\label{ntreeformat}

\begin{definition}[Variable dependency graph]
The variable dependency graph of a set $S$ of premises is a directed graph, with the set of variables as vertices, and with as edges the set

$$\{(x,y)|\textrm{there is a }t\xrightarrow{U}t'\textrm{ in }S\textrm{ such that }x\textrm{ occurs in }t\textrm{ and }y\textrm{ in }t'\}$$

$S$ is well-founded if any backward chain of edges in its variable dependency graph is finite.

A pomset transition rule $\rho$ is pure if its set of premises is well-founded and moreover each variable in the rule occurs in the source or as the right-hand side of a positive premise.
\end{definition}

\begin{definition}[Ntree format]
A pomset transition rule $\rho$ is an ntree rule if it satisfies the following three criteria:

\begin{enumerate}
  \item $\rho$ is panth.
  \item $\rho$ is pure.
  \item The left-hand sides of positive premises in $\rho$ are single variables.
\end{enumerate}

A PTSS is in ntree format if it consists of ntree rules only.
\end{definition}

\begin{theorem}
For each PTSS $T$ in panth format, there exists a PTSS $T'$ in ntree format such that for any closed pomset transition rule $\frac{N}{\alpha}$ where $N$ contains only negative literals, $T\vdash\frac{N}{\alpha}\Leftrightarrow T'\vdash\frac{N}{\alpha}$.
\end{theorem}

\subsection{De Simone Format}\label{desimoneformat}

A De Simone language consists of a signature together with a PTSS whose pomset transition rules are in De Simone format, extended with pomset transition rules for recursion.

\begin{definition}[De Simone format]
Let $\Sigma$ be a signature. A pomset transition rule $\rho$ is in De Simone format if it has the form:

$$\frac{\{x_i\xrightarrow{U_i}y_i|i\in I\}}{f(x_1,\cdots,x_{ar(f)})\xrightarrow{U}t}$$

where $I\subseteq\{1,\cdots,ar(f)\}$ and the variables $x_i$ and $y_i$ are all distinct and the only variables that occur in $\rho$. Moreover, the target $t\in\mathbb{T}(\Sigma)$ does not contain variables $x_i$ for $i\in I$ and has no multiple occurrences of variables. We say that $f$ is the type and $U$ the action pomset of $\rho$.
\end{definition}

Assume a countably infinite set $\mathsf{RVar}$ of recursion variables, ranged over by $X,Y$. The recursive term over $\Sigma$ are given by the BNF grammar:

$$t::=X|f(t_1,\cdots,t_{ar(f)})|\fix(X=t)$$

where $X$ is any recursive variable, $f$ any function symbol in $\Sigma$, $\fix$ a binding construct, $t[u/X]$ denotes the recursive term $t$ in which each occurrence of $X$ is replaced by $u$ with $u$ a recursive term.

For every recursive term $\fix(X=t)$ and pomset $U\in\mathsf{Pom}$, its pomset transition rule is as follows:

$$\frac{t[\fix(X=t)/X]\xrightarrow{U}y}{\fix(X=t)\xrightarrow{U}y}$$

\begin{theorem}
If a PTSS is in De Simone format, then pomset trace equivalence is a congruence w.r.t. the PLTS associated with it.
\end{theorem}

On the expressiveness of De Simone languages for true concurrency, like the results of expressiveness of traditional De Simone languages \cite{SOS}, between which and its associated LTSs, and which and traditional process algebras such as SCCS, Meije, $aprACP_R$, $aprACP_F$, $aprACP^{r.e.}_{R}$, $aprACP^{p.e.}_{R}$, $aprACP_{U}$, CSP, etc, we believe that we can get the similar expressiveness results between De Simone languages for true concurrency and its associated PLTSs, and between De Simone languages for true concurrency and the corresponding truly concurrent process algebras. Because some related concrete truly concurrent process algebras are still absent, let the work on expressiveness of De Simone languages for true concurrency be a future one.

\subsection{GSOS Format}\label{gsosformat}

\begin{definition}[GSOS format]
A pomset transition rule $\rho$ is in GSOS format if it has the form

$$\frac{\{x_i\xrightarrow{U_{ij}}y_{ij}|1\leq i\leq ar(f),1\leq j\leq m_i\}\cup\{x_i\xnrightarrow{V_{ik}}|1\leq i\leq ar(f),1\leq k\leq n_i\}}{f(x_1,\cdots,x_{ar(f)})\xrightarrow{W}t}$$

where $m_i,n_i\in\mathbb{N}$, and the variables $x_i$ and $y_{ij}$ are all distinct and the only variables that occur in $\rho$, $U_{ij},V_{ik},W\in\mathsf{Pom}$.

A (finitary) GSOS language is a finite set of GSOS rules over a finite signature, a finite set $\mathsf{Act}$ of actions and a finite set $\mathsf{Pom}$ of pomsets over $\mathsf{Act}$.
\end{definition}

In comparison to De Simone rules in the above section, GSOS rules allow for negative premises, as well as for multiple occurrences of variables in left-hand sides of premises and in the target.

Each GSOS language allows a stratification, and is therefore ws-complete and positive after reduction.

A GSOS language may contain junk rules that support no transitions in the associated PLTS. Also, whether a pomset transition rule in a GSOS language is junk is decidable. So, junk rules can be removed from a GSOS language without altering the associated PLTS.

Simply replacing traditional single action transition by pomset transition, it is reasonable to construct a GSOS term $U2CM_n$ that behaves as a universal 2-counter machine on input $n$. Then $U2CM_n\sim_{p}U^{\omega}$, $U2CM_n\sim_{s}U^{\omega}$, $U2CM_n\sim_{hp}U^{\omega}$ and $U2CM_n\sim_{hhp}U^{\omega}$, where $U^{\omega}\xrightarrow{U}U^{\omega}$, if and only if the 2-counter machine diverges on input $n$. Therefore, GSOS languages are Turing powerful, despite their finiteness restrictions.

A finite GSOS language can be extended to an infinitary one by allowing a countable set of GSOS rules over a countable signature, a countable set $\mathsf{Act}$ of actions and a countable set $\mathsf{Pom}$ of pomsets over $\mathsf{Act}$. By imposing more restrictions on an infinitary GSOS language, such as the so-called boundedness and simplicity, its associated PLTS is regular, i.e., it is finitely branching and the set of closed terms reachable from any closed term is finite.

An axiomatization provides a set of axioms to reason whether a term is equivalent to another, usually equipped with a PTSS, and the soundness and completeness bridging between them. Similarly to the work on automation of axiomatization modulo bisimulation equivalence, it is also reasonable to provide an algorithm to compute a sound and complete axiomatization modulo truly concurrent bisimulation equivalences, i.e., two closed terms can be equated by the axiom system if and only if they are truly concurrent bisimilar in the associated PLTS.

A GSOS language can be extended to a recursive one by adding recursive term same to that of recursive De Simone language in the above section. Let $\mathsf{CREC}(\Sigma)$ denote the set of recursive terms that do not contain free recursive variables for some signature $\Sigma$. By introducing a special inert constant $\Omega$ into a recursive GSOS language and a divergence predicate $\uparrow$ into a PLTS, for every recursive GSOS language with $\Omega$, there exists a least sound and supported PLTS with $\uparrow$ over $\mathsf{CREC}(\Sigma)$. Intuitively, $\Omega$ is akin to the constant $\mathbf{0}$ in Milner's CCS without any transitions, and for a term $t$, $t\uparrow$ if its initial transitions are not fully specified, which is usually caused either when the initial behaviour of $t$ depends on under-specified arguments like $\Omega$, or in the presence of unguarded recursive definitions.

\subsection{RBB Safe Format}\label{rbbformat}

Now, we extend the silent step $\tau$ into $\mathsf{Act}$.

\begin{definition}[Branching pomset, step bisimulation]\label{BPSB}
Assume a special termination predicate $\downarrow$, and let $\surd$ represent a state with $\surd\downarrow$. Let $\mathcal{P}$ be a PLTS. A branching pomset bisimulation is a relation $R\subseteq\mathcal{C}(\mathcal{P})\times\mathcal{C}(\mathcal{P})$, such that:
 \begin{enumerate}
   \item If $(\mathbf{C}_1,\mathbf{C}_2)\in R$, and $\mathbf{C}_1\xrightarrow{U}\mathbf{C}_1'$ then
   \begin{itemize}
     \item either $U\equiv \tau^*$, and $(\mathbf{C}_1',\mathbf{C}_2)\in R$;
     \item or there is a sequence of (zero or more) $\tau$-transitions $\mathbf{C}_2\xrightarrow{\tau^*} \mathbf{C}_2^0$, such that $(\mathbf{C}_1,\mathbf{C}_2^0)\in R$ and $\mathbf{C}_2^0\xRightarrow{U}\mathbf{C}_2'$ with $(\mathbf{C}_1',\mathbf{C}_2')\in R$.
   \end{itemize}
   \item If $(\mathbf{C}_1,\mathbf{C}_2)\in R$, and $\mathbf{C}_2\xrightarrow{U}\mathbf{C}_2'$ then
   \begin{itemize}
     \item either $X\equiv \tau^*$, and $(\mathbf{C}_1,\mathbf{C}_2')\in R$;
     \item or there is a sequence of (zero or more) $\tau$-transitions $\mathbf{C}_1\xrightarrow{\tau^*} \mathbf{C}_1^0$, such that $(\mathbf{C}_1^0,\mathbf{C}_2)\in R$ and $\mathbf{C}_1^0\xRightarrow{U}\mathbf{C}_1'$ with $(\mathbf{C}_1',\mathbf{C}_2')\in R$.
   \end{itemize}
   \item If $(\mathbf{C}_1,\mathbf{C}_2)\in R$ and $\mathbf{C}_1\downarrow$, then there is a sequence of (zero or more) $\tau$-transitions $\mathbf{C}_2\xrightarrow{\tau^*}\mathbf{C}_2^0$ such that $(\mathbf{C}_1,\mathbf{C}_2^0)\in R$ and $\mathbf{C}_2^0\downarrow$.
   \item If $(\mathbf{C}_1,\mathbf{C}_2)\in R$ and $\mathbf{C}_2\downarrow$, then there is a sequence of (zero or more) $\tau$-transitions $\mathbf{C}_1\xrightarrow{\tau^*}\mathbf{C}_1^0$ such that $(\mathbf{C}_1^0,\mathbf{C}_2)\in R$ and $\mathbf{C}_1^0\downarrow$.
 \end{enumerate}

We say that $\mathbf{C}_1$, $\mathbf{C}_2$ are branching pomset bisimilar, written $\mathbf{C}_1\approx_{bp}\mathbf{C}_2$, if there exists a branching pomset bisimulation $R$, such that $(\emptyset,\emptyset)\in R$.

By replacing pomset transitions with steps, we can get the definition of branching step bisimulation. When $\mathbf{C}_1$ and $\mathbf{C}_2$ are branching step bisimilar, we write $\mathbf{C}_1\approx_{bs}\mathbf{C}_2$.
\end{definition}

\begin{definition}[Rooted branching pomset, step bisimulation]\label{RBPSB}
Assume a special termination predicate $\downarrow$, and let $\surd$ represent a state with $\surd\downarrow$. Let $\mathcal{P}$ be a PLTS. A rooted branching pomset bisimulation is a relation $R\subseteq\mathcal{C}(\mathcal{P})\times\mathcal{C}(\mathcal{P})$, such that:
 \begin{enumerate}
   \item If $(\mathbf{C}_1,\mathbf{C}_2)\in R$, and $\mathbf{C}_1\xrightarrow{U}\mathbf{C}_1'$ then $\mathbf{C}_2\xrightarrow{U}\mathbf{C}_2'$ with $\mathbf{C}_1'\approx_{bp}\mathbf{C}_2'$.
   \item If $(\mathbf{C}_1,\mathbf{C}_2)\in R$, and $\mathbf{C}_2\xrightarrow{U}\mathbf{C}_2'$ then $\mathbf{C}_1\xrightarrow{U}\mathbf{C}_1'$ with $\mathbf{C}_1'\approx_{bp}\mathbf{C}_2'$.
   \item If $(\mathbf{C}_1,\mathbf{C}_2)\in R$ and $\mathbf{C}_1\downarrow$, then $\mathbf{C}_2\downarrow$.
   \item If $(\mathbf{C}_1,\mathbf{C}_2)\in R$ and $\mathbf{C}_2\downarrow$, then $\mathbf{C}_1\downarrow$.
 \end{enumerate}

We say that $\mathbf{C}_1$, $\mathbf{C}_2$ are rooted branching pomset bisimilar, written $\mathbf{C}_1\approx_{rbp}\mathbf{C}_2$, if there exists a rooted branching pomset bisimulation $R$, such that $(\emptyset,\emptyset)\in R$.

By replacing pomset transitions with steps, we can get the definition of rooted branching step bisimulation. When $\mathbf{C}_1$ and $\mathbf{C}_2$ are rooted branching step bisimilar, we write $\mathbf{C}_1\approx_{rbs}\mathbf{C}_2$.
\end{definition}

\begin{definition}[Branching (hereditary) history-preserving bisimulation]\label{BHHPB}
Assume a special termination predicate $\downarrow$, and let $\surd$ represent a state with $\surd\downarrow$. Let $\mathcal{P}$ be a PLTS. A branching history-preserving (hp-) bisimulation is a posetal relation $R\subseteq\mathcal{C}(\mathcal{P})\overline{\times}\mathcal{C}(\mathcal{P})$ such that:

 \begin{enumerate}
   \item If $(\mathbf{C}_1,f,\mathbf{C}_2)\in R$, and $\mathbf{C}_1\xrightarrow{a_1}\mathbf{C}_1'$ then
   \begin{itemize}
     \item either $a_1\equiv \tau$, and $(\mathbf{C}_1',f[a_1\mapsto \tau],\mathbf{C}_2)\in R$;
     \item or there is a sequence of (zero or more) $\tau$-transitions $\mathbf{C}_2\xrightarrow{\tau^*} \mathbf{C}_2^0$, such that $(\mathbf{C}_1,f,\mathbf{C}_2^0)\in R$ and $\mathbf{C}_2^0\xrightarrow{a_2}\mathbf{C}_2'$ with $(\mathbf{C}_1',f[a_1\mapsto a_2],\mathbf{C}_2')\in R$.
   \end{itemize}
   \item If $(\mathbf{C}_1,f,\mathbf{C}_2)\in R$, and $\mathbf{C}_2\xrightarrow{a_2}\mathbf{C}_2'$ then
   \begin{itemize}
     \item either $a_2\equiv \tau$, and $(\mathbf{C}_1,f[a_2\mapsto \tau],\mathbf{C}_2')\in R$;
     \item or there is a sequence of (zero or more) $\tau$-transitions $\mathbf{C}_1\xrightarrow{\tau^*} \mathbf{C}_1^0$, such that $(\mathbf{C}_1^0,f,\mathbf{C}_2)\in R$ and $\mathbf{C}_1^0\xrightarrow{a_1}\mathbf{C}_1'$ with $(\mathbf{C}_1',f[a_2\mapsto a_1],\mathbf{C}_2')\in R$.
   \end{itemize}
   \item If $(\mathbf{C}_1,f,\mathbf{C}_2)\in R$ and $\mathbf{C}_1\downarrow$, then there is a sequence of (zero or more) $\tau$-transitions $\mathbf{C}_2\xrightarrow{\tau^*}\mathbf{C}_2^0$ such that $(\mathbf{C}_1,f,\mathbf{C}_2^0)\in R$ and $\mathbf{C}_2^0\downarrow$.
   \item If $(\mathbf{C}_1,f,\mathbf{C}_2)\in R$ and $\mathbf{C}_2\downarrow$, then there is a sequence of (zero or more) $\tau$-transitions $\mathbf{C}_1\xrightarrow{\tau^*}\mathbf{C}_1^0$ such that $(\mathbf{C}_1^0,f,\mathbf{C}_2)\in R$ and $\mathbf{C}_1^0\downarrow$.
 \end{enumerate}

$\mathbf{C}_1,\mathbf{C}_2$ are branching history-preserving (hp-)bisimilar and are written $\mathbf{C}_1\approx_{bhp}\mathbf{C}_2$ if there exists a branching hp-bisimulation $R$ such that $(\emptyset,\emptyset,\emptyset)\in R$.

A branching hereditary history-preserving (hhp-)bisimulation is a downward closed branching hp-bisimulation. $\mathbf{C}_1,\mathbf{C}_2$ are branching hereditary history-preserving (hhp-)bisimilar and are written $\mathbf{C}_1\approx_{bhhp}\mathbf{C}_2$.
\end{definition}

\begin{definition}[Rooted branching (hereditary) history-preserving bisimulation]\label{RBHHPB}
Assume a special termination predicate $\downarrow$, and let $\surd$ represent a state with $\surd\downarrow$. Let $\mathcal{P}$ be a PLTS. A rooted branching history-preserving (hp-) bisimulation is a weakly posetal relation $R\subseteq\mathcal{C}(\mathcal{P})\overline{\times}\mathcal{C}(\mathcal{P})$ such that:

 \begin{enumerate}
   \item If $(\mathbf{C}_1,f,\mathbf{C}_2)\in R$, and $\mathbf{C}_1\xrightarrow{a}\mathbf{C}_1'$, then $\mathbf{C}_2\xrightarrow{a}\mathbf{C}_2'$ with $\mathbf{C}_1'\approx_{bhp}\mathbf{C}_2'$.
   \item If $(\mathbf{C}_1,f,\mathbf{C}_2)\in R$, and $\mathbf{C}_2\xrightarrow{a}\mathbf{C}_2'$, then $\mathbf{C}_1\xrightarrow{a}\mathbf{C}_1'$ with $\mathbf{C}_1'\approx_{bhp}\mathbf{C}_2'$.
   \item If $(\mathbf{C}_1,f,\mathbf{C}_2)\in R$ and $\mathbf{C}_1\downarrow$, then $\mathbf{C}_2\downarrow$.
   \item If $(\mathbf{C}_1,f,\mathbf{C}_2)\in R$ and $\mathbf{C}_2\downarrow$, then $\mathbf{C}_1\downarrow$.
 \end{enumerate}

$\mathbf{C}_1,\mathbf{C}_2$ are rooted branching history-preserving (hp-)bisimilar and are written $\mathbf{C}_1\approx_{rbhp}\mathbf{C}_2$ if there exists a rooted branching hp-bisimulation $R$ such that $(\emptyset,\emptyset,\emptyset)\in R$.

A rooted branching hereditary history-preserving (hhp-)bisimulation is a downward closed rooted branching hp-bisimulation. $\mathbf{C}_1,\mathbf{C}_2$ are rooted branching hereditary history-preserving (hhp-)bisimilar and are written $\mathbf{C}_1\approx_{rbhhp}\mathbf{C}_2$.
\end{definition}

Let $[]$ be the context symbol and $C[]$ denote a context which is a term with one occurrence of $[]$.

\begin{definition}[Wild-nested context]
The set of wild-nested contexts over a signature $\Sigma$ is defined inductively by:

\begin{enumerate}
  \item $[]$ is wild-nested.
  \item If $C[]$ is wild-nested, and the ith argument of function symbol $f\in\Sigma$ is wild-nested, then \\
  $f(t_1,\cdots,t_{i-1},C[],t_{i+1},\cdots,t_{ar(f)})$ is wild-nested.
\end{enumerate}
\end{definition}

\begin{definition}[Patience rule]
A patience rule for the ith argument of a function symbol $f\in\Sigma$ is a GSOS rule of the form

$$\frac{x_i\xrightarrow{\tau}y}{f(x_1,\cdots,x_{ar(f)})\xrightarrow{\tau}f(x_1,\cdots,x_{i-1},y,x_{i+1},\cdots,x_{ar(f)})}$$
\end{definition}

Assume that every argument of each function symbol is labelled either tame or wild.

\begin{definition}[RBB safe format]
A PTSS $T$ in panth format is RBB safe if there exists a tame/wild labelling of arguments of function symbols such that each of its pomset transition rules $\rho$ is:

\begin{enumerate}
  \item Either a patience rule for a wild argument of a function symbol.
  \item Or a rule with source $f(x_1,\cdots,x_{ar(f)})$ for which the following requirements are fulfilled:
  \begin{itemize}
    \item right-hand sides of positive premises do not occur in left-hand sides of premises of $\rho$;
    \item if ith argument of $f$ is wild and does not have a patience rule in $T$, then $x_i$ does not occur in left-hand sides of premises of $\rho$;
    \item if ith argument of $f$ is wild and has a patience rule in $T$, then $x_i$ occurs no more than once in the left-hand sides of a premise of $\rho$, where this premise
    \begin{itemize}
      \item is positive,
      \item does not contain the relation symbol $\xrightarrow{\tau}$,
      \item has left-hand side $x_i$;
    \end{itemize}
    \item right-hand sides of positive premises and variables $x_i$ for $i$ a wild argument of $f$ only occur at wild-nested positions in the target of $\rho$.
  \end{itemize}
\end{enumerate}
\end{definition}

\begin{theorem}
If a ws-complete PTSS is RBB safe, then the rooted branching pomset bisimulation equivalence that it induces is a congruence.
\end{theorem}

\begin{theorem}
If a ws-complete PTSS is RBB safe, then the rooted branching step bisimulation equivalence that it induces is a congruence.
\end{theorem}

\begin{theorem}
If a ws-complete PTSS is RBB safe, then the rooted branching hp-bisimulation equivalence that it induces is a congruence.
\end{theorem}

\begin{theorem}
If a ws-complete PTSS is RBB safe, then the rooted branching hhp-bisimulation equivalence that it induces is a congruence.
\end{theorem}

\subsection{Precongruence Formats for Truly Concurrent Behavioural Preorders}\label{precongruence}

\begin{theorem}
If a PTSS is in path format, then the pomset simulation preorder that it induces is a precongruence.
\end{theorem}

\begin{theorem}
If a PTSS is in path format, then the step simulation preorder that it induces is a precongruence.
\end{theorem}

\begin{theorem}
If a PTSS is in path format, then the hp-simulation preorder that it induces is a precongruence.
\end{theorem}

\begin{theorem}
If a PTSS is in path format, then the hhp-simulation preorder that it induces is a precongruence.
\end{theorem}

\begin{definition}[Ready simulation format]
A panth rule is in ready simulation format if the variables at the right-hand sides of its positive premises do not occur in the left-hand sides of its premises.

A PTSS is in ready simulation format if all its transition rules are.
\end{definition}

\begin{theorem}
If a ws-complete PTSS is in ready simulation format, then the ready pomset simulation preorder that it induces is a precongruence.
\end{theorem}

\begin{theorem}
If a ws-complete PTSS is in ready simulation format, then the ready step simulation preorder that it induces is a precongruence.
\end{theorem}

\begin{theorem}
If a ws-complete PTSS is in ready simulation format, then the ready hp-simulation preorder that it induces is a precongruence.
\end{theorem}

\begin{theorem}
If a ws-complete PTSS is in ready simulation format, then the ready hhp-simulation preorder that it induces is a precongruence.
\end{theorem}

\begin{definition}[$\mathcal{F}$-winterized]
A GSOS language $T$ is $\mathcal{F}$-winterized if there exists a tame/wild labelling of arguments of function symbols such that for each of its transition rule $\rho$, with source $f(x_1,\cdots,x_{ar(f)})$, the following requirements are fulfilled:

\begin{itemize}
  \item Right-hand sides of positive premises do not occur in left-hand sides of premises of $\rho$.
  \item If the ith argument of $f$ is wild and there is a positive premise $x_i\xrightarrow{U}y$ in $\rho$ where $y$ occurs in the target, then this is the only premise in $\rho$ to have $x_i$ as its left-hand side.
  \item Right-hand sides of positive premises and variables $x_i$ for $i$ a wild argument of $f$ only occur at wild-nested positions in the target of $\rho$.
  \item The target of $\rho$ has no multiple occurrences of variables.
\end{itemize}
\end{definition}

\begin{theorem}
If a PTSS is $\mathcal{F}$-winterized, then the readies preorder that it induces is a precongruence.
\end{theorem}

\begin{definition}[Ready pomset trace format]
A panth rule $\rho$ is in ready pomset trace format if:

\begin{enumerate}
  \item Right-hand sides of positive premises do not occur in left-hand sides of premises of $\rho$.
  \item Each pair variables that occur at distinct positions in the target of $\rho$ are not connected in the symmetric closure of the variable dependency graph of the premises of $\rho$.
\end{enumerate}

A PTSS is in ready pomset trace format if all its pomset transition rules are.
\end{definition}

\begin{theorem}
If a ws-complete PTSS is in ready pomset trace format, then the ready pomset trace preorder that it induces is a precongruence.
\end{theorem}

\begin{theorem}
The failures preorder that a De Simone language induces is a precongruence.
\end{theorem}

\begin{definition}[L cool]
A PTSS in path format is L cool if there exists a tame/wild labelling of arguments of function symbols such that each of its pomset transition rules $\rho$ satisfies the following restrictions:

\begin{enumerate}
  \item Each variable in $\rho$ that occurs in a wild argument of the source, or as the right-hand side of a premise, occurs exactly once either as the left-hand side of a premise or at a wild-nested position in the target of $\rho$.
  \item The variable dependency graph of the set of premises of $\rho$ does not contain an infinite forward chain of edges.
\end{enumerate}
\end{definition}

\begin{theorem}
If a PTSS is in L cool format, then the accepting pomset trace preorder that it induces is a precongruence.
\end{theorem}

\begin{definition}[Pomset trace format]
A finite PTSS in tyft format is in pomset trace format if each pomset transition rule satisfies the following restrictions:

\begin{enumerate}
  \item It contains finite many premises.
  \item Each variable occurs either as the right-hand side of a premise or in the source.
  \item No variable occurs more than once in the left-hand sides of the premises and in the target.
\end{enumerate}
\end{definition}

\begin{theorem}
If a PTSS is in pomset trace format, then the pomset trace equivalence that it induces is congruence.
\end{theorem}

\subsection{Trace Congruences}\label{tracec} 

\begin{definition}[Completed pomset trace congruence]
Let $\mathcal{F}$ be some rule format and $T_0$ a PTSS in $\mathcal{F}$ format over a signature $\Sigma_0$. Two closed term $t$ and $u$ are completed pomset trace congruent w.r.t. $T_0$ and $\mathcal{F}$, denoted $t\sim^{\mathcal{F}}_{ct}u$, if for every PTSS $T_1$ in $\mathcal{F}$ format over some signature $\Sigma_1$ that can be added in an operationally conservative fashion to $T_0$, and for every context $C[]$ over $\Sigma_0\oplus\Sigma_1$, the PLTS associated with $T_0\oplus T_1$ yields $C[t]\sim_{ct}C[u]$.
\end{definition}

\begin{definition}[Denial formula]
The sets of denial formulae over $\mathsf{Act}$ $\mathcal{D}_{p}$, $\mathcal{D}_{s}$, $\mathcal{D}_{hp}$ and $\mathcal{D}_{hhp}$ which are sublanguages of the modal logics PL, SL, HPL and BCL in \cref{bcl}, are give by the following BNF grammars:

$$\varphi_{p}::=\mathbf{true}|\varphi_{p_1}\wedge\varphi_{p_2}|\semangle{\mathbf{a},\overline{\mathbf{b}}<c}\varphi_{p}|\neg\langle a\rangle\mathbf{true}$$
$$\varphi_{s}::=\mathbf{true}|\varphi_{s_1}\wedge\varphi_{s_2}|(\semangle{a_1}\otimes\cdots\otimes\semangle{a_n}\varphi_{s}|\neg\langle a\rangle\mathbf{true}$$
$$\varphi_{hp}::=\mathbf{true}|\varphi_{hp_1}\wedge\varphi_{hp_2}|\semangle{\mathbf{a},\overline{\mathbf{b}}<c}\varphi_{hp}|\neg\langle a\rangle\mathbf{true}$$
$$\varphi_{hhp}::=\mathbf{true}|\varphi_{hhp_1}\wedge\varphi_{hhp_2}|(\mathbf{a},\overline{\mathbf{b}}<c)\varphi_{hhp}|\langle a\rangle \varphi_{hhp}|\neg\langle a\rangle\mathbf{true}$$

where $a,c\in\mathsf{Act}$, $\mathbf{a},\mathbf{b}$ are tuples of actions, a state satisfies $\neg\langle a\rangle\mathbf{true}$ if and only if $a$ is not one of its initial actions.
\end{definition}

For a PLTS $\mathcal{P}$ and $\mathbf{C},\mathbf{C}'\in\mathcal{C}(\mathcal{P})$, we define the corresponding equivalence relations as follows:

\begin{enumerate}
  \item $\mathbf{C}\sim_{\mathcal{D}_{p}}\mathbf{C}'$ if and only if for all denial formulae $\varphi_{p}$, $(\mathcal{P},\mathbf{C})\models\varphi_{p}\Leftrightarrow (\mathcal{P},\mathbf{C}')\models\varphi_{p}$.
  \item $\mathbf{C}\sim_{\mathcal{D}_{s}}\mathbf{C}'$ if and only if for all denial formulae $\varphi_{s}$, $(\mathcal{P},\mathbf{C})\models\varphi_{s}\Leftrightarrow (\mathcal{P},\mathbf{C}')\models\varphi_{s}$.
  \item $\mathbf{C}\sim_{\mathcal{D}_{hp}}\mathbf{C}'$ if and only if for all denial formulae $\varphi_{hp}$, $(\mathcal{P},\mathbf{C})\models\varphi_{hp}\Leftrightarrow (\mathcal{P},\mathbf{C}')\models\varphi_{hp}$.
  \item $\mathbf{C}\sim_{\mathcal{D}_{hhp}}\mathbf{C}'$ if and only if for all denial formulae $\varphi_{hhp}$, $(\mathcal{P},\mathbf{C})\models\varphi_{hhp}\Leftrightarrow (\mathcal{P},\mathbf{C}')\models\varphi_{hhp}$.
\end{enumerate}

\begin{theorem}
Let $T$ be a GSOS language. The equivalence relation $\sim_{\mathcal{D}_{p}}$ induced by the PLTS associated with $T$ coincides with the completed pomset trace congruence $\sim^{GSOS}_{ct}$ w.r.t. $T$ and the GSOS format.
\end{theorem}

\begin{theorem}
In every finitely branching PLTS, two configurations are ready pomset bisimulation equivalent if and only if they satisfy exactly the same denial formulae $\varphi_{p}$.
\end{theorem}

\begin{theorem}
Assume a stratifiable PTSS in pure ntyft/ntyxt format containing at least one constant in its signature. Then, for every pair of closed terms $t,u$,

$$t\sim^{pure~ntyft/ntyxt}_{ct}u\Leftrightarrow t\sim_{PL}u$$
\end{theorem} 
\newpage\section{Many-Sorted Higher-Order Languages}\label{mshol}

In many-sorted higher-order languages, a sort is a fundamental syntactic and semantic category that partitions the domain of discourse into disjoint, non-empty collections of entities, including individuals, functions, predicates, and higher-order constructs. Each term, variable, function symbol, and predicate in such a language is explicitly assigned a unique sort or a sequence of sorts, enforcing strict type discipline and enabling well-formed constraints that prevent ill-typed compositions in both syntax and model-theoretic interpretation. Higher-order indicates that the logical and linguistic framework supports not only quantification and abstraction over basic individuals of base sorts but also over functions, predicates, and even higher-order functions and predicates that themselves take or return entities of other sorts, thereby enabling the expression of complex structural properties and higher-level abstractions within a strictly sorted type system.

Under the background of many-sorted higher-order language, for a PTSS, variable binding mechanisms and substitutions are unavoidable. Variables are classed into formal variables and actual variables: traditionally, formal variables are the variable names declared in the definition of a function, procedure, or lambda abstraction, which serve as placeholders that specify the number, order, and type (or sort) of inputs the construct expects, without referring to concrete values; while actual parameters (also called arguments) are the concrete expressions, values, or variables supplied to a function or procedure at the point of its application or invocation, they are substituted for the corresponding formal parameters during the evaluation or execution of the function. A pomset transition rule can be deemed as a procedure for a pomset transition relation by substituting actual terms for the formal variables. We use $x,y,z$ as actual variables and $x^*,y^*,z^*$ as formal variables. For a substitution $\sigma$, $y[z/x]$ is a syntactic construct called a substitution harness, while $\sigma(y)[\sigma(z)/x]$ is the evaluated term after the application of $\sigma$.

In this chapter, in \cref{aw}, we introduce the actual terms and we introduce the formal terms in \cref{fw}. We introduce the actual and formal pomset transition rules in \cref{aftr} and carry the operational conservative extension to a many-sorted higher-order setting in \cref{oce2}.

\subsection{The Actual World}\label{aw}

Let $\mathsf{Var}$ be a denumerable set of actual variables ranged over by $x,y,z$ with a set of sorts, and $\overrightarrow{O}$ be a sequence of $O_1\cdots O_k$ and $\overrightarrow{O_i}$ a sequence of $O_{i1}\cdots O_{ik}$, where $i,k\in\mathbb{N}$.

\begin{definition}[Many-sorted higher-order signature]
A many-sorted higher-order signature $\Sigma$ is a set of function symbols:

$$f:\overrightarrow{S_1}.S_1\times\cdots\times\overrightarrow{S_{ar(f)}}.S_{ar(f)}\rightarrow S$$

where $S_{ij}$, $S_i$ and $S$ are sorts.
\end{definition}

\begin{definition}[Actual terms]
Let $\Sigma$ be a many-sorted higher-order signature. The collection $\mathbb{A}(\Sigma)$ of actual terms over $\Sigma$ is the least set satisfying:

\begin{itemize}
  \item If $x\in \mathsf{Var}$, then $x\in\mathbb{A}(\Sigma)$.
  \item For every $f:\overrightarrow{S_1}.S_1\times\cdots\times\overrightarrow{S_{ar(f)}}.S_{ar(f)}\rightarrow S$, $f(\overrightarrow{x_1}.t_1\times\cdots\times\overrightarrow{x_{ar(f)}}.t_{ar(f)})$ is an actual term of sort $S$, if:
      \begin{itemize}
        \item Every $t_i$ is an actual term of sort $S_i$.
        \item Every $\overrightarrow{x_i}$ consists of distinct actual variables in $\mathsf{Var}$ of sorts $\overrightarrow{S_i}$.
      \end{itemize}
\end{itemize}
\end{definition}

\begin{definition}[Closed actual terms]
Free actual variables in actual terms are defined inductively as follows:

\begin{itemize}
  \item Every $x\in \mathsf{Var}$ occurs freely in $x$.
  \item If $x$ occurs freely in $t_i$ and does not occur freely in $\overrightarrow{x_i}$, then $x$ occurs freely in $f(\overrightarrow{x_1}.t_1\times\cdots\times\overrightarrow{x_{ar(f)}}.t_{ar(f)})$.
\end{itemize}

An actual term is closed if it does not contain any free actual variables.
\end{definition}

\begin{definition}[Actual substitutions]
An actual substitution is a sort preserving map $\sigma:\mathsf{Var}\rightarrow\mathbb{A}(\Sigma)$, while $x$ and $\sigma(x)$ have the same sort. $\sigma$ can be extended to $\sigma:\mathbb{A}(\Sigma)\rightarrow\mathbb{A}(\Sigma)$, and $\sigma(t)$ for $t\in\mathbb{A}(\Sigma)$ is obtained by replacing each free variable $x$ by $\sigma(x)$. The postfix $\_[t/x]$ maps $x$ to $t$ and is inert otherwise. As usual, actual terms are considered modulo $\alpha$-conversion under an actual substitution. 
\end{definition}

\subsection{The Formal World}\label{fw}

Assume a many-sorted higher-order signature $\Sigma$. The set of formal variables is defined as $\mathsf{Var}^*=\{x^*|x\in \mathsf{Var}\}$, where $x^*$ and $x$ are of the same sort. A formal term $t^*$ is an actual term with possible occurrences of formal variables and substitution harnesses.

\begin{definition}[Formal terms]
The collection $\mathbb{F}(\Sigma)$ for formal terms over a many-sorted higher-order signature $\Sigma$ is the least set satisfying:

\begin{itemize}
  \item If $x\in \mathsf{Var}$, then $x\in\mathbb{F}(\Sigma)$.
  \item If $x\in \mathsf{Var}^*$, then $x\in\mathbb{F}(\Sigma)$.
  \item For every $f:\overrightarrow{S_1}.S_1\times\cdots\times\overrightarrow{S_{ar(f)}}.S_{ar(f)}\rightarrow S$, $f(\overrightarrow{x_1}.t_1^*\times\cdots\times\overrightarrow{x_{ar(f)}}.t_{ar(f)}^*)$ is a formal term of sort $S$, if:
      \begin{itemize}
        \item Every $t_i^*$ is a formal term of sort $S_i$.
        \item Every $\overrightarrow{x_i}$ consists of distinct actual variables in $\mathsf{Var}$ of sorts $\overrightarrow{S_i}$.
      \end{itemize}
  \item If $t^*$ and $u^*$ are formal terms of sorts $S_0$ and $S_1$ respectively, and $x\in \mathsf{Var}$ is of sort $S_1$, then $t^*[u^*/x]$ is a formal term of sort $S_0$.
\end{itemize}
\end{definition}

\begin{definition}[Formal substitutions]
A formal substitution is a sort preserving map $\sigma^*:\mathsf{Var}^*\rightarrow\mathbb{A}(\Sigma)$, while $x^*$ and $\sigma^*(x^*)$ have the same sort. $\sigma^*$ can be extended to $\sigma^*:\mathbb{F}(\Sigma)\rightarrow\mathbb{A}(\Sigma)$, and $\sigma^*(t^*)$ for $t^*\in\mathbb{F}(\Sigma)$ is obtained by: (1) replacing each formal variable $x^*$ by $\sigma^*(x^*)$; (2) the substitution harnesses in $t^*$ become postfix one and the result is an actual term.
\end{definition}

\subsection{Actual and Formal Pomset Transition Rules}\label{aftr}

\begin{definition}[Actual pomset transition rules]
An actual pomset transition rule is an expression of the form $\frac{H}{\alpha}$, where $H$ is a set of literals and $\alpha$ is a positive literal.
\end{definition}

\begin{definition}[Formal pomset transition rules]
A formal pomset transition rule is an expression of the form $\frac{H^*}{\alpha^*}$, where:

\begin{itemize}
  \item $H^*$ is a set of premises of the form $t^*\xrightarrow{U}t^{*'}$, or $t^* P$, or $t^*\xnrightarrow{U}$, or $t^*\neg P$.
  \item $\alpha^*$ is the conclusion of the form $t^*\xrightarrow{U}t^{*'}$ or $t^* P$.
\end{itemize}

Where $t^*, t^{*'}\in\mathbb{F}(\Sigma)$, $U\in\mathsf{Pom}$, $P\in\mathsf{Pred}$. A higher-order PTSS is a set of formal pomset transition rules.
\end{definition}

\subsection{Operational Conservative Extension}\label{oce2} 

$FV(t^*)$ denotes the set of formal variables that occur in the formal term $t^*$.

\begin{definition}[$FV(t^*)$]
The set $FV(t^*)$ is defined inductively by:

$$FV(x^*)\triangleq x^*$$
$$FV(f(\overrightarrow{x_1}.t_1^*\times\cdots\times\overrightarrow{x_{ar(f)}}.t_{ar(f)}^*))\triangleq FV(t_1^*)\cup\cdots\cup FV(t_{ar(f)}^*)$$
$$FV(t^*[u^*/x])\triangleq FV(t^*)\cup FV(u^*)$$
\end{definition}

$EV(t^*)$ denotes a more restricted set of formal variables that occur in the formal term $t^*$.

\begin{definition}[$EV(t^*)$]
The set $FV(t^*)$ is defined inductively by:

$$EV(x^*)\triangleq x^*$$
$$EV(f(\overrightarrow{x_1}.t_1^*\times\cdots\times\overrightarrow{x_{ar(f)}}.t_{ar(f)}^*))\triangleq EV(t_1^*)\cup\cdots\cup EV(t_{ar(f)}^*)$$
$$EV(t^*[u^*/x])\triangleq EV(t^*)$$
\end{definition}

\begin{definition}[Source dependency]
For a formal pomset transition rule $\rho^*$, the formal variables in $\rho^*$ that are source-dependent, are defined inductively by:

\begin{enumerate}
  \item If $t^*$ is the source of $\rho^*$, then all formal variables in $EV(t^*)$ are source-dependent in $\rho^*$.
  \item If $t^*\xrightarrow{U}t^{*'}$ is a premise of $\rho^*$, and all formal variables in $FV(t^*)$ are source-dependent in $\rho^*$, then all formal variables in $EV(t^{*'})$ are source-dependent in $\rho^*$.
\end{enumerate}

A formal pomset transition rule $\rho^*$ is source-dependent if all formal variables in $FV(\rho^*)$ are source-dependent.
\end{definition}

\begin{theorem}[Operational conservative extension]
Let $T_0$ and $T_1$ be higher-order PTSSs over many-sorted higher-order signatures $\Sigma_0$ and $\Sigma_0\oplus\Sigma_1$, respectively. Under the following conditions, $T_0\oplus T_1$ is an operational conservative extension of $T_0$.

\begin{enumerate}
  \item Each $\rho^*\in T_0$ is source-dependent.
  \item For each $\rho^*\in T_1$,
  \begin{itemize}
    \item either the source of $\rho^*$ is fresh;
    \item or $\rho^*$ has a premise of the form $t_0^*\xrightarrow{U}t_1^*$ or $t_0^* P$, where
    \begin{itemize}
      \item $t_0^*\in\mathbb{F}(\Sigma_0)$;
      \item $FV(t_0^*)\subseteq EV(u^*)$, where $u^*$ denotes the source of $\rho^*$;
      \item $t_1^*$, every $a\in U$, $P$ are fresh.
    \end{itemize}
  \end{itemize}
\end{enumerate}
\end{theorem}
\newpage\section{Denotational Semantics}\label{ds}

The theory of processes concerns operational semantics, axiomatic semantics and denotational semantics, which describe different aspects of behaviours of processes. This thesis stresses on operation semantics of processes, and some atomic semantics in \cref{aoa}. Most meta-theory of processes concerns the relationship between the operational semantics and the axiomatic semantics, and there exists a good correspondence between them: it is often possible to automatically translate an operational theory of processes into an axiomatic one; moreover, in certain circumstances, it is also possible to derive an SOS semantics from an axiomatic one.

Denotational semantics of processes can provide not only a thorough insight on process theory, but also a complementary benefit. We wish the denotational semantics matches the operational semantics of processes, if exactly, it is called a fully abstract one.

In this chapter, we give the fully abstract denotational semantics for a truly concurrent process algebra. We introudce the $\Sigma$-domains and truly concurrent prebisimulations in \cref{sigmad} and \cref{tcp}. Finally, we give the denotational semantics of a truly concurrent process algebra in \cref{dsatcpa}.

\subsection{$\Sigma$-Domains}\label{sigmad}

In Scott-Strachey approach \cite{DS} of denotational semantics, domains are the key concepts. In this chapter, we introduce the related concepts of $\Sigma$-domain, which are coming from algebraic semantics \cite{IACA} \cite{AlgebraicSemantics} \cite{TS}.

\begin{definition}[$\Sigma$-domains]
Let $\Sigma$ be a signature which contains a distinguished constant $\Omega$. A $\Sigma$-domain consists of $\langle A, \leq_A, \Sigma_A\rangle$, where:

\begin{enumerate}
  \item $A$ is the carrier set.
  \item $\leq_A$ is a partial order over $A$.
  \item $\langle A, \leq_A\rangle$ is an algebraic complete partial order (cpo) \cite{TS}.
  \item $\Sigma_A$ is a set of continuous functions $\{f_A:A^{ar(f)}\rightarrow A|f\in\Sigma\}$ with respect to $\leq_A$.
  \item $\Omega_A$ is $\bot_A$ which is the least element with respect to $\leq_A$.
\end{enumerate}

Sometimes, we use $A$ to denote $\langle A, \leq_A, \Sigma_A\rangle$.
\end{definition}

\begin{definition}[$\Sigma$-domain homomorphism]
A mapping $\varphi:A\rightarrow B$ between two $\Sigma$-domains $\langle A,\leq_A,\Sigma_A\rangle$ and $\langle B,\leq_B,\Sigma_B\rangle$ is a $\Sigma$-domain homomorphism if:

\begin{enumerate}
  \item for every $f\in\Sigma$, it holds that $\varphi(f_A(a_1,\cdots,a_{ar(f)}))=f_B(\varphi(a_1),\cdots,\varphi(a_{ar(f)}))$, where $a_1,\cdots,a_{ar(f)}\in A$.
  \item it is continuous with respect to $\leq$.
\end{enumerate}

A $\Sigma$-domain homomorphism is $\Sigma$-domain isomorphic if it is has an inverse.

Sometimes, we write $\Sigma$-homomorphism for $\Sigma$-domain homomorphism and $\Sigma$-isomorphism for $\Sigma$-domain isomorphism.
\end{definition}

We recall that $T(\Sigma)$ denotes the set of closed terms over $\Sigma$, $\mathsf{CREC}(\Sigma)$ to denote the set of closed terms over $\Sigma$ that may contain occurrences from a countably infinite set of $\mathsf{RVar}$ of recursive variables ranged over by $X,Y$.

The interpretation $A\sembrack{-}$ of $\mathsf{CREC}(\Sigma)$ is a $\Sigma$-algebra $A$ associates each term in $\mathsf{CREC}(\Sigma)$ with a mapping from substitutions to $A$, which is defined inductively as follows, where $\sigma$ is any mapping from recursive variables to $A$.

$$A\sembrack{X}\sigma\triangleq \sigma(X)$$
$$A\sembrack{f(t_1,\cdots,t_{ar(f)})}\sigma\triangleq f_A(A\sembrack{t_1}\sigma,\cdots,A\sembrack{t_{ar(f)}}\sigma)$$
$$A\sembrack{\fix(X=t)}\sigma\triangleq Y.\lambda d.A\sembrack{t}\sigma'$$

where $Y$ denotes the least fixpoint operator, $d$ is a metavariable ranging over $A$, $\sigma'(X)\triangleq d$, and $\sigma'(Y)\triangleq\sigma(Y)$ for $Y\neq X$.

For any recursive term $t$ there is a sequence of finite approximations $t_n\in\mathsf{CREC}(\Sigma)$ for $n\in\mathbb{N}$, such that for any $\Sigma$-domain $A$, $$A\sembrack{t}=\bigvee_{n\in\mathbb{N}}A\sembrack{t_n}$$

\subsection{Truly Concurrent Prebisimulations}\label{tcp}

Let us recall the definition of PLTS in \cref{pomlts}.

\pomsetlts*

For $n\geq 0$ and $s=U_1\cdots U_n\in\mathsf{Pom}^*$, then $p\xrightarrow{s}q$ if and only if there exist processes $p_0,\cdots,p_n$ such that $p=p_0\xrightarrow{U_1}p_1\xrightarrow{U_2}\cdots\xrightarrow{U_{n-1}}p_{n-1}\xrightarrow{U_n}p_n=q$. For $p\in\mathsf{Proc}$ and $U\in\mathsf{Pom}$, we define:

$$\mathsf{initials}(p)\triangleq\{U\in\mathsf{Pom}|\exists q:p\xrightarrow{U}q\}$$
$$\mathsf{sort}(p)\triangleq\{U\in\mathsf{Pom}|\exists s\in\mathsf{Pom}^*,p',p''\in\mathsf{Proc}:p\xrightarrow{s}p'\xrightarrow{U}p''\}$$
$$\mathsf{derivatives}(p,U)\triangleq\{q|p\xrightarrow{U}q\}$$

A PLTS is sort-finite if and only if $\mathsf{sort}(p)$ is finite for every $p\in\mathsf{Proc}$.

\begin{definition}[Pomset synchronization tree]
The set of countably branching pomset synchronization trees over $\mathsf{Pom}$, denoted $\mathbf{PST}_{\infty}(\mathsf{Pom})$, is the set of infinitary terms generated by the following inductive definition:

$$\frac{\{U_i\in\mathsf{Pom},t_i\in\mathbf{PST}_{\infty}(\mathsf{Pom})\}_{i\in I}}{\sum_{i\in I}U_i:t_i[+\Omega]\in\mathbf{PST}_{\infty}(\mathsf{Pom})}$$

where $I$ is a countable index set, $0=\sum_{i\in\emptyset}U_i:t_i$ which stands for the one-node synchronization tree, a representation of an inactive process, $\Omega=\sum_{i\in\emptyset}U_i:t_i+\Omega$ which stands that the behaviours of the synchronization tree are completely unspecified, $[+\Omega]$ means optional inclusion of $\Omega$ as a summand.
\end{definition} 

Let $\mathbf{PST}(\mathsf{Pom})$ denote the finite synchronization tree, i.e., the above set of terms is built using only finite summations. The set of synchronization trees $\mathbf{PST}_{\infty}(\mathsf{Pom})$ can be turned into a PLTS with divergence by stipulating that, for $t\in\mathbf{PST}_{\infty}(\mathsf{Pom})$:

\begin{itemize}
  \item $t\uparrow$ if and only if $\Omega$ is a summand of $t$.
  \item $t\xrightarrow{U_i}t_i$ if and only if $U_i:t_i$ is a summand of $t$.
\end{itemize} 

In the following, we give the definitions of pomset prebisimulation.

\begin{definition}[Pomset prebisimulation]\label{PSPB}
Let $\langle\mathsf{Proc},\mathsf{Act},\{\xrightarrow{U}|U\in\mathsf{Pom}\},\mathsf{Pred}\rangle$ be a PLTS with divergence $\uparrow$ and $p,q\in\mathsf{Proc}$. Let $\mathsf{Rel}(\mathsf{Proc})$ denote the set of binary relations over $\mathsf{Proc}$. Define the functional $F:\mathsf{Rel}(\mathsf{Proc})\rightarrow\mathsf{Rel}(\mathsf{Proc})$, such that:

\begin{align*}
\begin{aligned}
F(R)=&\{(p,q)|\forall U\in\mathsf{Pom}\\
  &\bullet p\xrightarrow{U}p'\mbox{ implies } q\xrightarrow{U}q', \mbox{ and }(p',q')\in R\\
  &\bullet p\downarrow\mbox{ implies }\bigg(q\downarrow\mbox{ and }\Big(q\xrightarrow{U}q'\mbox{ implies } p\xrightarrow{U}p',\mbox{ and }(p',q')\in R\Big)\bigg)\}\\
\end{aligned}
\end{align*}

A relation $R$ is a pomset prebisimulation if and only if $R\subseteq F(R)$. We say that $p$, $q$ are pomset prebisimilar, written $p\prepomset q$, if there exists a pomset prebisimulation $R$, such that $(p,q)\in R$.
\end{definition}

By use of the functional $F$ in \cref{PSPB}, we can obtain a pomset behavioural preorder which applies itself inductively as follows:

$$\prepomsets{0}\triangleq\mathsf{Proc}\times\mathsf{Proc}$$
$$\prepomsets{n+1}\triangleq F(\prepomsets{n})$$
$$\prepomsets{\omega}\triangleq\bigcap_{n\geq 0}\prepomsets{n}$$

\begin{definition}[Finitary pomset prebisimulation]
Let $\langle\mathsf{Proc},\mathsf{Act},\{\xrightarrow{U}|U\in\mathsf{Pom}\},\mathsf{Pred}\rangle$ be a PLTS and $p,q\in\mathsf{Proc}$. The finitary pomset prebisimulation $\prepomset^F$ over $\mathsf{Proc}$ is defined as follows: $p\prepomset^F q$ if and only if, for every $t\in\mathbf{PST}(\mathsf{Pom})$, $t\prepomset p$ implies $t\prepomset q$.
\end{definition} 

They are obvious that $\prepomset~\subseteq~\prepomsets{\omega}~\subseteq~\prepomset^F$. Moreover, the inclusions are strict for infinitely branching PLTSs, and collapse to equalities for finitely branching PLTSs. 

Let $\langle\mathsf{Proc},\mathsf{Act},\{\xrightarrow{U}|U\in\mathsf{Pom}\},\mathsf{Pred}\rangle$ be a PLTS and $p,q\in\mathsf{Proc}$. For every $P\subseteq\mathsf{Pom}$, we define the following relations.

$\{\pretesting^{P}_{(P,n)}|n\geq 0\}$

$p\pretesting^{P}_{(P,0)}q\mbox{ if and only if } \mathrm{true}$
\begin{align*}
\begin{aligned}
&p\pretesting^{P}_{(P,n+1)}q\mbox{ if and only if }\\
  &\bullet \forall U\in P,p\xrightarrow{U}p'\mbox{ implies } q\xrightarrow{U}q', \mbox{ and }p'\pretesting^{P}_{(P,n)}q'\\
  &\bullet \mathsf{initials(p)}\subseteq P\mbox{ and }p\downarrow\mbox{ imply }\\
  &\bigg(\mathsf{initials(q)}\subseteq P\mbox,q\downarrow,\Big(\forall U\in P, q\xrightarrow{U}q'\mbox{ implies } p\xrightarrow{U}p',\mbox{ and }p'\pretesting^{P}_{(P,n)}q'\Big)\bigg)\\
\end{aligned}
\end{align*}

\begin{proposition}
For every $n\geq 0$ and $P\subseteq\mathsf{Pom}$, the following statements hold:

\begin{enumerate}
  \item $\pretesting^{P}_{(P,n)}$ is a preorders.
  \item For $p,q\in\mathsf{Proc}$, $p\pretesting^{P}_{(P,n+1)}q\mbox{ implies } p\pretesting^{P}_{(P,n)}q$. 
  \item Assume that $P\subseteq P'\subseteq\mathsf{Pom}$, then for $p,q\in\mathsf{Proc}$, $p\pretesting^{P'}_{(P,n)}q\mbox{ implies } p\pretesting^{P}_{(P,n)}q$.
\end{enumerate}
\end{proposition}

We define:

$$\pretesting^{P}_{(P,\omega)}\triangleq\bigcap_{n\geq 0}\pretesting^{P}_{(P,n)}$$
$$p\pretesting^{fin}_{(P,\omega)}q\triangleq\forall P\subseteq_{fin}\mathsf{Pom}.p\pretesting^{P}_{(P,\omega)}q$$

where $P\subseteq_{fin}\mathsf{Pom}$ means that $P$ is a finite subset of $\mathsf{Pom}$.

\begin{lemma}
For every $t\in\mathbf{PST}(\mathsf{Pom})$ and $p\in\mathsf{Proc}$, it holds that: $t\prepomset p\mbox{ if and only if } t\pretesting^{fin}_{(P,\omega)}p$.
\end{lemma}

\begin{lemma}
The preorder $\pretesting^{fin}_{(P,\omega)}$ is finitary.
\end{lemma}

\begin{theorem}
For $p,q\in\mathsf{Proc}$ in any transition system, $p\prepomset^{F} q\mbox{ if and only if } p\pretesting^{fin}_{(P,\omega)}q$.
\end{theorem}

\begin{corollary}
For $p,q\in\mathsf{Proc}$ in any sort-finite transition system, $p\prepomset^{F} q\mbox{ if and only if } p\prepomsets{\omega}q$.
\end{corollary} 

\subsection{Denotational Semantics of A Truly Concurrent Process Algebra}\label{dsatcpa}

\begin{definition}[The signature]
The signature $\Sigma$ consists of:

\begin{enumerate}
  \item A set of atomic actions $\mathsf{Act}$ ranged over $a,b,\cdots$.
  \item A set of pomsets $\mathsf{Pom}$ over $\mathsf{Act}$ ranged over $U,V,\cdots$.
  \item A constant $\delta$ denoting inaction without any behaviour.
  \item A constant $\epsilon$ denoting empty action which terminates immediately and successfully.
  \item A distinguished constant $\Omega$.
  \item The communication action $\gamma(a,b)$.
  \item The binary operator $\cdot$ as the sequential composition, i.e., for processes $p$ and $p'$, the process $p\cdot p'$ firstly executes $p$ followed $p'$. The process $p\cdot p'$ is abbreviated as $pp'$.
  \item The binary operator $\sharp$ as the conflict composition, i.e., the process $a\sharp b$ either executes $a$ and its successors or $b$ and its successors.
  \item The binary operator $+$ as the alternative composition, i.e., for processes $p$ and $p'$, the process $p+p'$ either executes $p$ or $p'$ alternatively.
  \item The binary operator $\between$ as the concurrent composition, i.e., for processes $p$ and $p'$, the process $p\between p'$ means $p$ and $p'$ execute concurrently, i.e., in parallel but may be with unstructured communications.
  \item The binary operator $\parallel$ as the parallel composition, i.e., for processes $p$ and $p'$, the process $p\parallel p'$ executes $p$ and $p'$ in parallel.
  \item The binary operator $\mid$ as the communication merge, i.e., for processes $p$ and $p'$, the process $p\mid p'$ executes with synchronous communications. 
  $$a\mid b=\begin{cases}
                \gamma(a,b), & a\leq^c b;\\
                \delta, & \mbox{otherwise}.
            \end{cases}$$
            where $a\leq^c b$ denotes that there exists a communication between $a$ and $b$.
  \item The unary operator $\Theta$ as confliction eliminator, i.e., for process $p$, the process $\Theta(p)$ eliminates and the $\sharp$ relations between actions in $p$.
  \item The binary unless operator $\triangleleft$ as an auxiliary operator to confliction eliminator $\Theta$.
  \item The unary operator $\partial_H$ as the encapsulation, i.e., for process $p$, the process $\partial_H(p)$ renames all actions of $p$ in the set $H$ to $\delta$.
  \item A set of recursive variables $\mathsf{RVar}$ ranged over $X,Y,\cdots$.
\end{enumerate}

Brackets are omitted whenever possible, with sequential composition $\cdot$ having a higher precedence than concurrent composition $\between$, parallel composition $\parallel$ and communication merge $\mid$. Concurrent composition $\between$, parallel composition $\parallel$ and communication merge $\mid$ have the same precedences which are higher than alternative composition $+$ and conflict composition $\sharp$, and alternative composition $+$ and conflict composition $\sharp$ have the same precedences. And we assume that $\fix(X=p)$ has the lowest precedence of all the operators in $\Sigma$.
\end{definition}

\begin{definition}[Syntax of basic process language]
The syntax of the basic process language $\mathbf{M}$ is given by the following BNF grammar:

$p::=\Omega~|~\delta~|~\epsilon~|~U~|~\gamma(a,b)~|~p\cdot p~|~a\sharp b~|~X~|~p+p~|~p\parallel p~|~p\mid p~|~p\between p~|~\Theta(p)~|~p\triangleleft p~|~\partial_H(p)~|~\fix(X=p)$

where $a,b\in\mathsf{Act}$, $U\in\mathbf{Pom}$, $p\in \mathsf{Proc}$, $X$ is recursive variable and $\fix(X=p)$ stands for the process defined by the recursive equation $X=p$.
\end{definition}

\subsubsection{Operational Semantics}\label{os9}

In this section, we give the operational semantics of the language $\mathbf{M}$. The predicate $\surd$ represents successful termination, $\xrightarrow{a}\surd$ represents successful termination after execution of the action $a\in\mathsf{Act}$, $\xrightarrow{U}\surd$ represents successful termination after execution of the action $U\in\mathsf{Pom}$ and $\xrightarrow{ }\surd$ represents successful termination without execution of the any action. The divergence predicate $p\uparrow$ represents that $p$ is divergent, i.e., $p$ has an infinite internal computation. While the convergence predicate $p\downarrow$ represents that $p$ is not divergent $p\nuparrow$, i.e., convergent, $p$ has no infinite internal computation. The following are the PTSS of the language $\mathbf{M}$, where $p,q\in\mathsf{Proc}$.

The PTSS of $\Omega$ is as follows.

$$\frac{}{\Omega\uparrow}$$
$$\frac{}{\Omega\xrightarrow{\tau}\Omega}$$

The PTSS of action $\epsilon$, $a\in\mathsf{Act}$ and $U\in\mathsf{Pom}$ is as follows. Note that, there is no any transition rules for $\delta$.

$$\frac{}{\delta\downarrow}\quad\frac{}{\epsilon\downarrow}\quad \frac{}{a\downarrow} \quad\frac{}{U\downarrow}$$
$$\frac{}{\epsilon\xrightarrow{ }\surd}\quad\frac{}{a\xrightarrow{a}\surd}\quad\frac{}{U\xrightarrow{U}\surd}$$

The PTSS of sequential composition is as follows.

$$\frac{p\downarrow\quad q\downarrow}{p\cdot q\downarrow}$$
$$\frac{p\xrightarrow{U}\surd}{p\cdot q\xrightarrow{U} q}\quad\frac{p\xrightarrow{U}p'}{p\cdot q\xrightarrow{U} p'\cdot q}$$

The PTSS of alternative composition is as follows.

$$\frac{q\downarrow\quad q\downarrow}{p+q\downarrow}$$
$$\frac{p\xrightarrow{U}\surd}{p+ q\xrightarrow{U}\surd} \quad\frac{p\xrightarrow{U}p'}{p+ q\xrightarrow{U}p'} \quad\frac{q\xrightarrow{U}\surd}{p+ q\xrightarrow{U}\surd} \quad\frac{q\xrightarrow{U}q'}{p+ q\xrightarrow{U}q'}$$

The PTSS of concurrent composition is as follows.

$$\frac{p\downarrow\quad q\downarrow}{p\between q\downarrow}$$
$$\frac{p\xrightarrow{a}\surd\quad q\xrightarrow{b}\surd}{p\between q\xrightarrow{\step{a,b}}\surd} \quad\frac{p\xrightarrow{a}p'\quad q\xrightarrow{b}\surd}{p\between q\xrightarrow{\step{a,b}}p'}$$
$$\frac{p\xrightarrow{a}\surd\quad q\xrightarrow{b}q'}{p\between q\xrightarrow{\step{a,b}}q'} \quad\frac{p\xrightarrow{a}p'\quad q\xrightarrow{b}q'}{p\between q\xrightarrow{\step{a,b}}p'\between q'}$$
$$\frac{p\xrightarrow{a}\surd\quad q\xrightarrow{b}\surd}{p\between q\xrightarrow{\gamma(a,b)}\surd} \quad\frac{p\xrightarrow{a}p'\quad q\xrightarrow{b}\surd}{p\between q\xrightarrow{\gamma(a,b)}p'}$$
$$\frac{p\xrightarrow{a}\surd\quad q\xrightarrow{b}q'}{p\between q\xrightarrow{\gamma(a,b)}q'} \quad\frac{p\xrightarrow{a}p'\quad q\xrightarrow{b}q'}{p\between q\xrightarrow{\gamma(a,b)}p'\between q'}$$

The PTSS of parallel composition is as follows.

$$\frac{p\downarrow\quad q\downarrow}{p\parallel q\downarrow}$$
$$\frac{p\xrightarrow{a}\surd\quad q\xrightarrow{b}\surd}{p\parallel q\xrightarrow{\step{a,b}}\surd} \quad\frac{p\xrightarrow{a}p'\quad q\xrightarrow{b}\surd}{p\parallel q\xrightarrow{\step{a,b}}p'}$$
$$\frac{p\xrightarrow{a}\surd\quad q\xrightarrow{b}q'}{p\parallel q\xrightarrow{\step{a,b}}q'} \quad\frac{p\xrightarrow{a}p'\quad q\xrightarrow{b}q'}{p\parallel q\xrightarrow{\step{a,b}}p'\between q'}$$

The PTSS of communication merge is as follows.

$$\frac{p\downarrow\quad q\downarrow}{p\mid q\downarrow}$$
$$\frac{p\xrightarrow{a}\surd\quad q\xrightarrow{b}\surd}{p\mid q\xrightarrow{\gamma(a,b)}\surd} \quad\frac{p\xrightarrow{a}p'\quad q\xrightarrow{b}\surd}{p\mid q\xrightarrow{\gamma(a,b)}p'}$$
$$\frac{p\xrightarrow{a}\surd\quad q\xrightarrow{b}q'}{p\mid q\xrightarrow{\gamma(a,b)}q'} \quad\frac{p\xrightarrow{a}p'\quad q\xrightarrow{b}q'}{p\mid q\xrightarrow{\gamma(a,b)}p'\between q'}$$

The PTSS of encapsulation operator is as follows.

$$\frac{p\downarrow}{\partial_H(p)\downarrow}$$
$$\frac{p\xrightarrow{a}\surd\quad a\notin H}{\partial_H(p)\xrightarrow{a}\surd}\quad\frac{p\xrightarrow{a}p'\quad a\notin H}{\partial_H(p)\xrightarrow{a}\partial_H(p')}$$

The PTSS of confliction, confliction eliminator and the auxiliary unless operator is as follows, where $\leq$ is the execution order.

$$\frac{p\downarrow}{\Theta(p)\downarrow}\quad\frac{p\downarrow\quad q\downarrow}{p\triangleleft q\downarrow}$$
$$\frac{p\xrightarrow{a}\surd\quad a\sharp b}{\Theta(p)\xrightarrow{a}\surd} \quad\frac{p\xrightarrow{b}\surd\quad a\sharp b}{\Theta(p)\xrightarrow{b}\surd}$$
$$\frac{p\xrightarrow{a}p'\quad a\sharp b}{\Theta(p)\xrightarrow{a}\Theta(p')} \quad\frac{p\xrightarrow{b}p'\quad a\sharp b}{\Theta(p)\xrightarrow{b}\Theta(p')}$$
$$\frac{p\xrightarrow{c}\surd \quad q\xnrightarrow{b}\quad a\sharp b\quad c\leq a}{p\triangleleft q\xrightarrow{c}\surd}
\quad\frac{p\xrightarrow{c}p' \quad q\xnrightarrow{b}\quad a\sharp b\quad c\leq a}{p\triangleleft q\xrightarrow{c}p'}$$
$$\frac{p\xrightarrow{a}\surd \quad q\xnrightarrow{b}\quad a\sharp b}{p\triangleleft q\xrightarrow{ }\surd}
\quad\frac{p\xrightarrow{a}p' \quad q\xnrightarrow{b}\quad a\sharp b}{p\triangleleft q\xrightarrow{ }p'}$$
$$\frac{p\xrightarrow{a}\surd \quad q\xnrightarrow{c}\quad a\sharp b\quad b\leq c}{p\triangleleft q\xrightarrow{ }\surd}
\quad\frac{p\xrightarrow{a}p' \quad q\xnrightarrow{c}\quad a\sharp b\quad b\leq c}{p\triangleleft q\xrightarrow{ }p'}$$
$$\frac{p\xrightarrow{c}\surd \quad q\xnrightarrow{b}\quad a\sharp b\quad a\leq c}{p\triangleleft q\xrightarrow{ }\surd}
\quad\frac{p\xrightarrow{c}p' \quad q\xnrightarrow{b}\quad a\sharp b\quad a\leq c}{p\triangleleft q\xrightarrow{ }p'}$$

The PTSS of recursion is as follows.

$$\frac{\fix(X=p[p/X])\downarrow}{\fix(X=p)\downarrow}$$
$$\frac{\fix(X=p[p/X])\xrightarrow{U}p'}{\fix(X=p)\xrightarrow{U}p'}$$

\subsubsection{Full Abstractness}

\begin{definition}[Partial order over $\mathbf{PST}(\mathsf{Pom})$]
A partial order $\leq_{\mathbf{PST}(\mathsf{Pom})}$ is defined over $\mathbf{PST}(\mathsf{Pom})$ as follows, for two trees $t_1,t_2\in\mathbf{PST}(\mathsf{Pom})$, $t_1\leq_{\mathbf{PST}(\mathsf{Pom})}t_2$, if:

\begin{enumerate}
  \item If $\langle U,t_1'\rangle\in t_1$, then $t_1'\leq_{\mathbf{PST}(\mathsf{Pom})}t_2'$ for some $\langle U, t_2'\rangle\in t_2$.
  \item If $\bot\in t_2$, then $\bot\in t_1$.
  \item If $\langle U,t_2'\rangle\in t_2$, then either $\bot\in t_1$ or $t_1'\leq_{\mathbf{PST}(\mathsf{Pom})}t_2'$ for some $\langle U, t_1'\rangle\in t_1$.
\end{enumerate}
\end{definition}

\begin{lemma}
$\langle \mathbf{PST}(\mathsf{Pom}),\leq_{\mathbf{PST}(\mathsf{Pom})}\rangle$ is an algebraic cpo.
\end{lemma}

Then we can define functions over $\mathbf{PST}(\mathsf{Pom})$ for every function symbol in $\Sigma$: $\delta_{\mathbf{PST}(\mathsf{Pom})}$, $\epsilon_{\mathbf{PST}(\mathsf{Pom})}$, $U_{\mathbf{PST}(\mathsf{Pom})}$, $\cdot_{\mathbf{PST}(\mathsf{Pom})}$, $+_{\mathbf{PST}(\mathsf{Pom})}$, $\parallel_{\mathbf{PST}(\mathsf{Pom})}$, $\gamma(a,b)_{\mathbf{PST}(\mathsf{Pom})}$, $\mid_{\mathbf{PST}(\mathsf{Pom})}$, $\between_{\mathbf{PST}(\mathsf{Pom})}$, $\partial_H(p)_{\mathbf{PST}(\mathsf{Pom})}$, $\sharp_{\mathbf{PST}(\mathsf{Pom})}$, $\Theta(p)_{\mathbf{PST}(\mathsf{Pom})}$, $\triangleleft_{\mathbf{PST}(\mathsf{Pom})}$. They are all continuous operations on the algebra cpo $\mathbf{PST}(\mathsf{Pom})$.

\begin{proposition}
$\langle \mathbf{PST}(\mathsf{Pom}),\leq_{\mathbf{PST}(\mathsf{Pom})},\Sigma_{\mathbf{PST}(\mathsf{Pom})}\rangle$ is a $\Sigma$-domain.
\end{proposition}

\begin{theorem}[Full Abstraction for $\mathbf{PST}(\mathsf{Pom})$]
If $p,q\in\mathbf{M}$, then $p\prepomset^F q$ if and only if $\mathbf{PST}(\mathsf{Pom})\sembrack{p}\leq_{\mathbf{PST}(\mathsf{Pom})}\mathbf{PST}(\mathsf{Pom})\sembrack{q}$.
\end{theorem}

\bibliographystyle{elsarticle-num}
\newpage\bibliography{Refs-SOS}

\end{document}